\documentclass[10pt,letterpaper]{article}

\usepackage{amsmath,amssymb}
\usepackage{breqn}

\usepackage{changepage}

\usepackage{textcomp,marvosym}

\usepackage{cite}

\usepackage{nameref,hyperref}

\usepackage[right]{lineno}

\usepackage[nopatch=eqnum]{microtype}
\DisableLigatures[f]{encoding = *, family = * }

\usepackage[table]{xcolor}

\usepackage{array}

\newcolumntype{+}{!{\vrule width 2pt}}

\newlength\savedwidth

\definecolor{Mygreen}{RGB}{50, 150, 50}

\definecolor{MPD}{rgb}{0.1010, 0.7056, 0.6014}
\newcommand{\mpd}[1]{{\color{MPD}#1}}

\newcommand\thickhline{\noalign{\global\savedwidth\arrayrulewidth\global\arrayrulewidth 2pt}%
\hline
\noalign{\global\arrayrulewidth\savedwidth}}

\raggedright
\usepackage[aboveskip=1pt,labelfont=bf,labelsep=period,justification=raggedright,singlelinecheck=off]{caption}

\makeatletter
\renewcommand{\@biblabel}[1]{\quad#1.}
\makeatother

\usepackage{lastpage,fancyhdr,graphicx}
\usepackage{epstopdf}
\def\eattif#1.tiff{#1}
\usepackage{epstopdf}
\AppendGraphicsExtensions{.tiff}

\graphicspath{{Figures/}} 

\def\eattif#1.tif{#1}
\usepackage{epstopdf}
\AppendGraphicsExtensions{.tif}

\usepackage{algorithm}
\usepackage{algpseudocode}

\usepackage{subcaption} 
\usepackage[all]{xy}
\usepackage{amsthm} 
\newtheorem{theorem}{Theorem}

\newtheorem{lemma}[theorem]{Lemma}

\newtheorem{proposition}[theorem]{Proposition}

\begin{document}
\vspace*{0.2in}

\begin{flushleft}
{\Large
\textbf\newline{A computational framework for optimal and Model Predictive Control of stochastic gene regulatory networks} 
}
\newline
\\
Hamza Faquir\textsuperscript{1},
Manuel Pájaro\textsuperscript{2,3},
Irene Otero-Muras\textsuperscript{1\textcurrency},
\\
\bigskip
\textbf{1} Computational Synthetic Biology Group, Institute for Integrative Systems Biology (CSIC-UV), Spanish National Research Council, 46980 Valencia, Spain.
\\
\textbf{2} 
Department of Mathematics, Escola Superior de Enxeñaría Informática, University of Vigo, Campus Ourense, 32004 Ourense, Spain
\\
\textbf{3} 
CITMAga, Santiago de Compostela 15782, Spain
\\
\bigskip


\textpilcrow Membership list can be found in the Acknowledgments section.

* irene.otero.muras@csic.es

\end{flushleft}
\section*{Abstract}
Engineering biology requires precise control of biomolecular circuits, and Cybergenetics is the field dedicated to achieving this goal. A significant challenge in developing controllers for cellular functions is designing systems that can effectively manage molecular noise. To address this, there has been increasing effort to develop model-based controllers for stochastic biomolecular systems, where a major difficulty lies in accurately solving the chemical master equation.\\ 
In this work we develop a framework for optimal and Model Predictive Control of stochastic gene regulatory networks with three key advantageous features: high computational efficiency, the capacity to control the overall probability density function enabling the fine-tuning of the cell population to obtain complex shapes and behaviors (including bimodality and other emergent properties), and the capacity to handle high levels of intrinsic molecular noise.  Our method exploits an efficient approximation of the Chemical Master Equation using Partial Integro-Differential Equations, which additionally enables the development of an effective adjoint-based optimization.\\
We illustrate the performance of the methods presented through two relevant studies in Synthetic Biology: shaping bimodal cell populations and tracking moving target distributions via inducible gene regulatory circuits.\\

\section*{Author summary}
Synthetic biology is already demonstrating its potential to provide groundbreaking solutions to real-world problems in sectors such as biomedicine, environmental sustainability, and bioproduction by reprogramming cells to execute novel functions. In many applications, precise control of cellular functions is essential for achieving desired outcomes. This has led to a growing effort at the intersection of synthetic biology and control theory, known as cybergenetics, which focuses on developing strategies to regulate cellular behavior dynamically and accurately. Despite significant advances, a major challenge remains the effective management of systems with high inherent molecular noise. In this work, we propose a Model Predictive Control (MPC) framework that integrates optimal control theory with Partial Integro-Differential Equation (PIDE) models and adjoint-based optimization techniques. Our approach provides a powerful tool for controlling biocircuits under noise,  ultimately aiming to achieve more reliable synthetic biological systems.


\section*{Introduction}

Cybergenetics is an emerging area of research at the intersection between control engineering and synthetic biology aiming to efficiently control biomolecular networks for cellular regulation  \cite{Hsiao:18, Khammash:19,  Pedone:21, Perrino:21, Ruolo:21, Barajas:22, Caringella:23}.  
Illustrative advances in the area in the last years include the antithetic controller\cite{Aoki:19},  and the effective balancing of genetic toggle switches using Proportional Integral control \cite{Lugagne:17}, Reinforcement Learning \cite{Guarino:20} or Adaptive Control \cite{Brancato:23}.
The closed-loop control of inducible gene regulatory circuits has been achieved {\it in vivo} using chemical inducers \cite{Lugagne:17, Fiore:16} or osmotic signals \cite{Uhlendorf:12} in microfluidic platforms, and light signals via optogenetic regulation \cite{Milias-Argeitis:16}. Increasing interest is also arising in control through electric signals \cite{Therrell:21}. Platforms for closed loop feedback control have been developed for various cell types including bacteria, yeast \cite{Fiore:16} and mammalian cells \cite{Frei:22, Khazim:21, Pedone:21}.

One of the main challenges in controlling gene regulatory systems is the intrinsic stochasticity of gene expression. While model-based control based on Ordinary Differential Equations (ODEs) can be effective for systems with near-deterministic \cite{Uhlendorf:12, Lugagne:17}, in many cases the dynamics of the circuit is significantly influenced by molecular noise. The dynamics of stochastic gene regulatory systems are described by the Chemical Master Equation, which is mathematically intractable for control purposes. In this work, we address this challenge by utilizing an efficient approximation of the Chemical Master Equation based on Partial Integro-Differential Equations (PIDEs) \cite{Pajaro:17}.

The development of PIDE models for multidimensional gene regulatory systems in 2017 \cite{Pajaro:17} opened a path for the development of new algorithms and methods for synthetic biology that effectively tackle the molecular noise inherent to gene expression\cite{Sequeiros:23}. The PIDE describes the time evolution of the probability density function of the protein levels of the gene regulatory system. Probability density function control has been previously applied in other engineering fields \cite{Ren:18}, mainly on processes modelled by Fokker Planck equations \cite{Gaviraghi:16}. Unlike the Fokker-Planck approximation associated with the Chemical Langevin Equation \cite{gillespie:2000}, the PIDE approximation, based on the assumption of protein bursting, enables the simulation of highly stochastic systems far from the thermodynamic limit. Notably, in contrast to the Fokker-Planck approach, the PIDE approximation also allows for the simulation of bimodal systems and other emergent behaviors.

The PIDE approximation has been validated elsewhere \cite{Pajaro:17, Pajaro:19} against the Gillespie algorithm, demonstrating its accuracy and reliability, provided that the degradation rate of the mRNA is faster than the degradation of the proteins, a common assumption in gene regulatory networks. Here, we leverage PIDE models to develop a generic framework for effective Model Predictive Control (MPC) of stochastic gene regulatory networks.  MPC is widely used in industry due to its predictive nature, as well as its ability to explicitly incorporate  state and input constraints providing a wider range of flexibility in the control design compared to other controllers \cite{Lars:11}. MPC of cell populations has been  previously addressed using deterministic models (see for example \cite{Milias-Argeitis:16, Uhlendorf:12}). Here, we leverage PIDE models for control, making possible to tackle noise by controlling the evolution of the probability density function governing the dynamics of the cell population.

The PIDE formulation allows for the explicit calculation of the gradient of the cost function, enabling the efficient determination of the optimal input using the nonlinear conjugate gradient method. We demonstrate the effectiveness of the methods through two case studies relevant to synthetic biology. In the first example, we showcase how MPC can induce bimodal behavior in a gene regulatory network at the population level, which is unachievable without control. In the second example, we apply the method to track a moving target distribution in an inducible gene regulatory system.

\section*{Methods}
\subsection*{PIDE modelling of stochastic Gene Regulatory Networks}
\label{sec:model}

\begin{figure}
    \centering
   \includegraphics[scale=0.5]{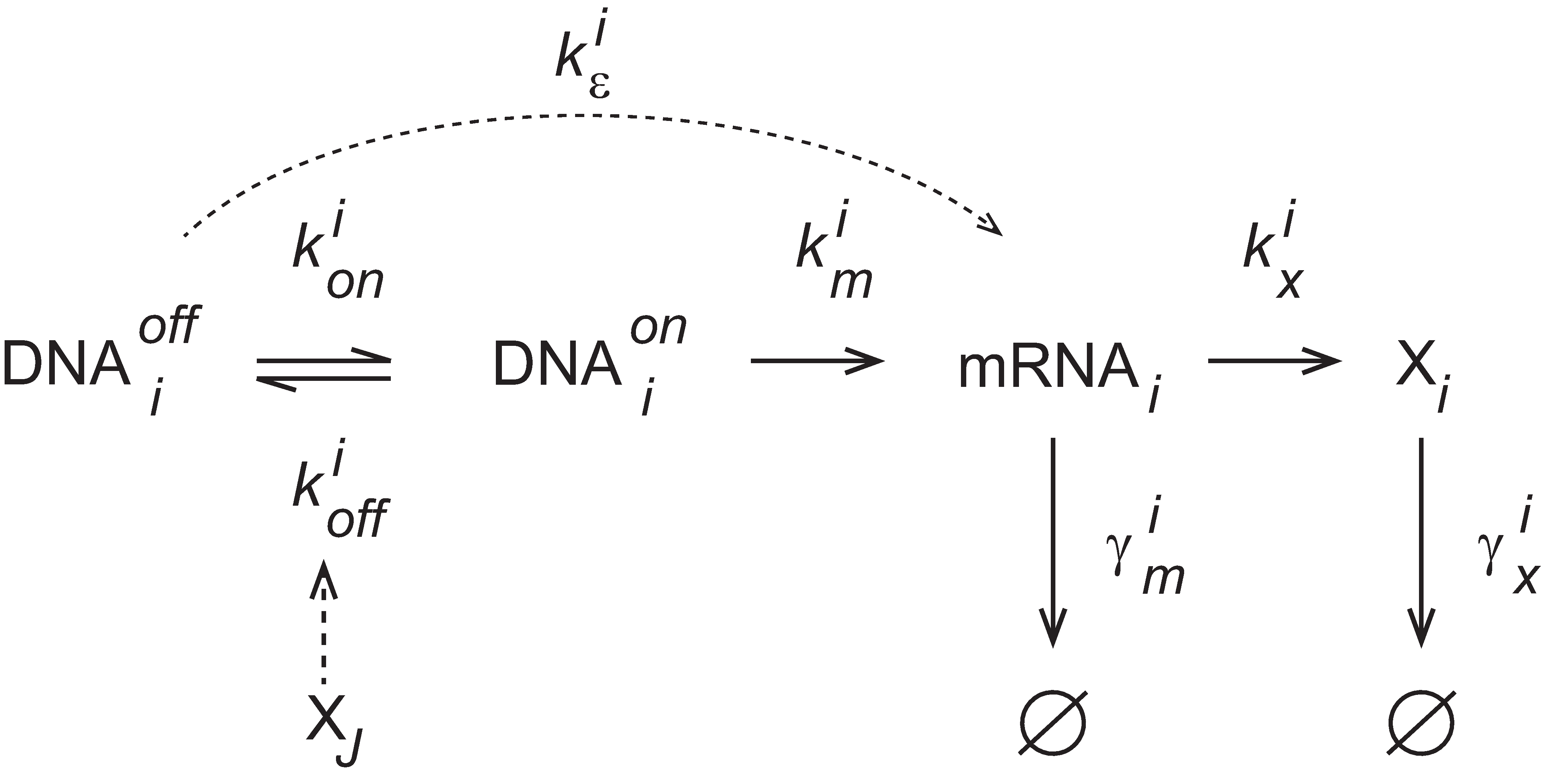}
    \caption{A schematic representation of a general gene regulatory network}
    \label{fig:network}
\end{figure}

We begin with a concise description of the Partial Integro-Differential Equation (PIDE) model for stochastic Gene Regulatory Networks introduced in \cite{Pajaro:17}.  A GRN consists of $n$ genes $\boldsymbol{G}=\left\{g_1, \ldots, g_n\right\}$ which transcribe  $n$ different messenger RNAs $\boldsymbol{M}=\left\{mRNA_1, \ldots, mRNA_n\right\}$. These mRNAs are subsequently translated into $n$ proteins $\boldsymbol{X}=\left\{X_1,  \ldots, X_n\right\}$, that can function either as activators or inhibitors of various genes in the network.  A schematic representation is depicted in Fig. \eqref{fig:network}.

Each promoter can switch from the inactive state $\left(DNAi_{\text {off }}\right)$ to the active state $\left(DNAi_{\mathrm{on}}\right)$ or vice versa with rate constants $k_{\mathrm{on}}^i$ and $k_{\mathrm{off}}^i$ respectively. Basal transcription can occur at a rate $\left(k_{\varepsilon}^i\right)$ which is normally lower than the transcription rate  $\left(k_m^i\right)$. Each type of messenger RNA $i$  is translated into the protein $X_i$ at rate constant $k_x^i$. Both messengers RNA and proteins are degraded with rates $\gamma_m^i$ and $\gamma_x^i(\mathbf{x})$ respectively.  Each gene $g_i$ can be regulated by any protein $X_i$ in the network (including cross and self-regulation), and the specific regulatory mechanisms are incorporated into the model through the function $c_i(\mathbf{x})$, where the $n$-vector $\mathbf{x}=\left(x_1, \ldots, x_n\right) \in \mathbb{R}_{+}^n$ is the amount of each protein type (examples of different input functions can be found in the Results Section. Under the mild assumption of protein bursting ($\gamma_m^i >>  \gamma_x^i$), which common in GRN since the protein lifetime is higher than that of messenger RNA, the temporal evolution of the probability density  function of the $n$ proteins $p: \mathbb{R}_{+} \times \mathbb{R}_{+}^n \rightarrow \mathbb{R}_{+}$ is given by the following equation:
\begin{eqnarray}
\dfrac{\partial p}{\partial t}(t, \mathbf{x})=  \sum_{i=1}^n  \Big(  & \dfrac{\partial}{\partial x_i}\left[\gamma_x^i(\mathbf{x}) x_i p(t,\mathbf{x})\right] -k_m^i c_i(\mathbf{x}) p(t,\mathbf{x}) \hspace{0.5cm} \nonumber \\
 & \displaystyle +k_m^i \int_0^{x_i} \omega_i\left(x_i-y_i\right) c_i\left(\mathbf{y}_i\right) p\left(t, \mathbf{y}_i\right) \mathrm{d} y_i\Big),
 \label{maineq}
\end{eqnarray}
where $\omega_i$ reads:
$$
\omega_i\left(x_i-y_i\right)=\frac{1}{b_i} \exp \left(-\frac{x_i-y_i}{b_i}\right)
$$
and $b_i=\frac{k_x^i}{\gamma_m^i}$ are dimensionless frequencies associated with translation which correspond with the mean protein amount produced per burst (burst size). Eq. \eqref{maineq}, denoted hereafter as GRN PIDE model, describes how the probability of having $\mathbf{x}$ number of proteins at time $t$ evolves. The model can be interpreted at the population level: if we consider an isogenic population of cells, Eq. \eqref{maineq} also describes how the frequency of a number of proteins $\mathbf{x}$ changes over time within a population. For a detailed derivation of the equation, we refer the reader to \cite{Pajaro:17}.

For simplicity of notation in Eq. \eqref{maineq}, we introduce the following terms:
\begin{align*}
&\Gamma(\mathbf{x})=(\gamma^1_x  x_1, \gamma^2_x x_2,...,\gamma^n_x x_n),\\
&I(p)(t,\mathbf{x})=\sum_{i=1}^{n}\left(k_m^i \int_0^{x_i} \! \omega_i(x_i-y_i)c_i(\mathbf{y}_i)p(t,\mathbf{y}_i) \, \mathrm{d}y_i\right).
\end{align*}

Then the initial value problem, can be written as follows
    \begin{align} \label{IVP}
	&\dfrac{\partial p}{\partial t}(t,\mathbf{x})  =  \nabla \cdot \left(\Gamma(\mathbf{x}) p(t,\mathbf{x})\right)+I(p)(t,\mathbf{x})- \sum_{i=1}^{n} k_m^i  \! c_i(\mathbf{x})p(t,\mathbf{x}),  \\
	&p(t_0,\mathbf{x})=p_0(\mathbf{x}), \nonumber
\end{align}
where $\nabla$ is the gradient operator with respect to $\mathbf{x}$.

In the following, we will define and focus the model on the domain $Q=I \times \Omega$, where $I$ is the time interval defined by  $I=\left[t_0, t_f\right]$, and $\Omega \subset \mathbb{R}_{+}^n$ is a bounded convex domain. In practice, this domain is rectangular $\displaystyle \Omega=\prod_{i=1}^n (0,a_i)$, with $a_i>0$ and boundary $\partial^+ \Omega$. Generally,  boundary consists of the faces of the hypercube given as domain (hypercubes of dimension $n-1$) that contain the vertex $(a_1,a_2,\ldots,a_n)$. By defining $\Sigma=I \times \partial^+ \Omega$,  we add to the initial value problem \eqref{IVP} the  Dirichlet boundary condition $p(t,\mathbf{x})=0, \ \forall (t,\mathbf{x}) \in \Sigma$. The initial boundary problem is then defined as follows:
\begin{equation}
  \mathcal{P}(p)=  \begin{cases} 
	&\dfrac{\partial p}{\partial t}(t,\mathbf{x})  =  \nabla \cdot (\Gamma p)+I(p)- \displaystyle \sum_{i=1}^{n} k_m^i  \! c_i(\mathbf{x})p(t,\mathbf{x}), \\
	&p(t_0,\mathbf{x})=p_0(\mathbf{x}), \\
 &p(t,\mathbf{x})=0, \; \; (t,\mathbf{x}) \in \Sigma.
 \end{cases}
 \label{IBP}
\end{equation}

Some relevant properties of this boundary value problem \cite{Canizo:19}  are included in the Appendix S1. 

\subsection*{Optimal control of gene regulation based on GRN PIDE models}
Next we introduce the algorithm for optimal control of stochastic gene regulatory networks based on GRN PIDE models, i.e. constrained by the expression \eqref{IBP}. The control variables are incorporated through the input functions \eqref{maineq} in $c_i:=c_i(\mathbf{x},\mathbf{u}), \ i=1,\dots,n$, where $\mathbf{u}(t)=(u_1(t),u_2(t),\dots, u_m(t))$ is the vector of time dependent controls. The control can be exerted through external stimuli (in the form of chemical species or light, for example). Provided that $c_i(.,\mathbf{u}),  \ i=1,\dots,n$ are smooth and bounded functions, the properties demonstrated in Appendix S1 are preserved. 

We define the set of admissible control space as:
\begin{equation}
U_{ad}=\{\mathbf{u} \in [L^2(I)]^m, \; \; \forall t \in I: u^{L}_j \leq u_j(t) \leq u^{U}_j\, \; \; \text{for } j=1,\dots m\} \subset [L^2(I)]^m.
\label{eq:AdmiSp}
\end{equation}
Equipped with the following norm:
$$
\|\mathbf{u}\|_2:=\sum_{j=1}^m\left( \int_I  |u_j(t)|^2 \mathrm{~d} t\right)^{\frac{1}{2}}. 
$$
This admissible input space represents the constraints on the control variables $\mathbf{u}$, where $\mathbf{u}^{L}$ and $\mathbf{u}^{U}$ are the lower and upper bounds of $\mathbf{u} \in \mathbb{R}^m$, with $u^{L}_j \leq u^{U}_j$ for $j=1\dots m$.

As discussed previously, the dynamics of a stochastic regulatory network are described by the evolution of the probability distribution function (PDF) of the system. Our optimal control goal is to drive the gene regulatory network dynamics described  by Eq. \eqref{IBP} to a target state represented by a reference  probability distribution $p_d$ at final time $t_f$. To this purpose we search for the control input profile $\mathbf{u}$ minimizing the difference between the solution of Eq. \eqref{IBP} and the reference state $p_d$ at final time $t_f$. The optimal control problem is formulated  as:
\begin{align}
	\min J(p, \mathbf{u}); \ J(p, \mathbf{u})&:=\frac{1}{2}\left\|p\left(\cdot, t_{f}\right)-p_{d}\left(\cdot\right)\right\|_{L^{2}(\Omega)}^{2}. \label{OC}
 \end{align}
 \begin{align}
	 \mathcal{P}(p,\mathbf{u})&:=  \begin{cases}
	& \displaystyle \dfrac{\partial p}{\partial t}(t,\mathbf{x})  =  \nabla \cdot (\Gamma p)+I(p,\mathbf{u})- \sum_{i=1}^{n} k_m^i  \! c_i(\mathbf{x},\mathbf{u})p(t,\mathbf{x}), \\
	&p(t_0,\mathbf{x})=p_0(\mathbf{x}), \\
 &p(t,\mathbf{x})=0, \; \; (t,\mathbf{x}) \in \Sigma .
 \end{cases}\label{mainp}
\end{align}
The difference between the two probability distributions is measured  by the $H:=L^2(\Omega)$ norm:
$$
\left\|p\left(\cdot, t_{f}\right)-p_{d}\left(\cdot\right)\right\|_{L^{2}(\Omega)}^{2}:= \int_{\Omega}|p\left(\mathbf{x}, t_{f}\right)-p_{d}\left(\mathbf{x}\right)|^2 \mathrm{d}\mathbf{x}.
$$
We define the following bounded Lipschitz domain $V$:
 \begin{align*}
     V &=\{ f \in H^1(\Omega) : f(x)=0 \text{ on } \partial^+ \Omega \},\\
     H^1(\Omega)&=\{ f \in L^2(\Omega): \partial_{x_i} f \in L^2(\Omega) \text{ for } i=1,....n \}.
 \end{align*}
and the Hilbert space:
\begin{equation} 
    F=\left\{f \in L^2(I ; V) \text{ such that } \partial_t f \in L^2(I ; V^*)\right\}.
\end{equation}
At this point, we note that the solution space of the initial boundary problem \eqref{IBP} belongs to the Hilbert space $F$. More details about these functional spaces and their inner products \cite{Troltzsh:10} are provided in Appendix S1. Under this formulation of the cost function, we impose constrains on the input \textbf{u} through the predefined bounds as stated in $U_{ad}$. A generic formulation of the cost function taking into account the input cost would be $\frac{1}{2}\left\|p\left(\cdot, t_{f}\right)-p_{d}\left(\cdot\right)\right\|_{L^{2}(\Omega)}^{2}+\lambda \left\| u \right\|_{L^{2}(I)}^{2}$, where $\lambda$ represent the weight of $\left\| u \right\|_{L^{2}(I)}^{2}$ on the cost function. For this formulation the analytical results still hold, and the term $\lambda \textbf{u}$ should be added to the gradient formulation.

The cost function $J$ depends explicitly both on $\mathbf{u}$ and $p$. Next, we define the control-to-state operator $\textit{S}: U_{ad}\rightarrow F$ that maps the control variable $\mathbf{u} \in U_{ad}$ to $S(\mathbf{u})=p(\mathbf{u}) \in F$, a solution to the initial boundary problem \eqref{mainp}. By introducing  the control-to-state operator in the cost function, we obtain the reduced cost function $\hat{J}$, which does not depend explicitly on $p$:
$$
\hat{J}(\mathbf{u})=J(S(\mathbf{u}),\mathbf{u}).
$$
The optimal control problem \eqref{OC} can be expressed as:
\begin{equation}
  \min_{\mathbf{u} \in U_{ad}} \hat{J}(\mathbf{u}), \label{uncon}  
\end{equation}
This transformation simplifies the derivation of mathematical results related to $\hat{J}$,  such us its gradient for deriving optimality conditions, as well as other mathematical properties. Next, we discuss some properties of the control-to-state operator $\textit{S}$  that will aid in the derivation of the first order optimality system.

\begin{proposition}
\label{diffcost}
    The control-to-state operator $\textit{S}:U_{ad}\rightarrow F $ is Fréchet differentiable, and the directional derivative in the direction $\mathbf{h}\in[L^2(I)]^m$ is given by
$$
\nabla_{\mathbf{u}}S(\mathbf{u})\cdot \mathbf{h}=g,
$$
where $g$ is  the weak solution to the initial-boundary value problem:
\begin{align} 
   &  \begin{array}{ll}
      \displaystyle \dfrac{\partial g}{\partial t}(t,\mathbf{x})  = &  \displaystyle \nabla \cdot (\Gamma g)+ I(g,\mathbf{u})-\sum_{i=1}^{n} k_m^i  \!c_i(\mathbf{x},\mathbf{u})g(t,\mathbf{x})+\nabla_{\mathbf{u}} I(p,\mathbf{u})\cdot \mathbf{h}(t)\\
       &  \displaystyle -\sum_{i=1}^{n} k_m^i  \! \nabla_{\mathbf{u}} c_i(\mathbf{x},\mathbf{u})\cdot \mathbf{h}(t)p(t,\mathbf{x}),  \nonumber
   \end{array} \\
 & g(t_0,\mathbf{x})=0, \label{Fr}\\
 & g(t,\mathbf{x})=0, \; \; (t,\mathbf{x}) \in \Sigma. \nonumber
\end{align}
\end{proposition}

The proof of Proposition \ref{diffcost} is included in Appendix S1.

This leads us to the main theorem of this section, which ensures the solvability of the unconstrained optimization problem.
\begin{theorem} \label{th:solUOP}
Let $p_0 \in V$, then there exist an optimal pair $(\Bar{p}, \Bar{\mathbf{u}})$ solution to the optimal control problem \eqref{OC}, such that $\Bar{p}=S(\Bar{\mathbf{u}})$ is the unique solution to \eqref{mainp} associated to $\Bar{\mathbf{u}}$,  and $\Bar{\mathbf{u}}$ is solution to the unconstrained optimization problem \eqref{uncon}.
\end{theorem}
The proof of Theorem  \ref{th:solUOP} is included in Appendix S1.
\subsubsection*{First order optimality conditions}
Here we derive the first-order optimality system that characterizes the solution to the optimal control problem \eqref{OC}. Due to Proposition \ref{diffcost} the operator $S$ is differentiable. Furthermore, the cost function $J$ is Fréchet differentiable and the reduced cost function $\hat{J}$ is differentiable. Since we will employ a gradient-based method to numerically solve the optimal control problem, we first derive the gradient of the function $\hat{J}$ on $\mathbf{u}$. Let $N(p)=\frac{1}{2}\|p-p_d\|_{L^2(\Omega)}^2$, we obtain:
\begin{equation} \label{eq:GradHatCF}
    \nabla \hat{J}=\left(\nabla_{\mathbf{u}}S(\mathbf{u})\right)^*N'(S(\mathbf{u})),
\end{equation}
where $\left(\nabla_{\mathbf{u}}S(\mathbf{u})\right)^*$ denotes the adjoint of the operator $\nabla_{\mathbf{u}}S(\mathbf{u})$. By taking the total derivative of $\mathcal{P}(S(\mathbf{u}),\mathbf{u})=0$ (which indicates that $S(\mathbf{u})$ and $\mathbf{u}$ are solutions of \eqref{mainp}), we get:
$$
\left(\nabla_{\mathbf{u}}S(\mathbf{u})\right)^*\left(\partial_p\mathcal{P}(S(\mathbf{u}),\mathbf{u})\right)^*+\left(\nabla_{\mathbf{u}}\mathcal{P}(S(\mathbf{u}),\mathbf{u})\right)^*=0,
$$
leading to the following expression 
\begin{equation}\label{eq:Su_star}
 \left(\nabla_{\mathbf{u}}S(\mathbf{u})\right)^*=-\left(\nabla_{\mathbf{u}}\mathcal{P}(S(\mathbf{u}),\mathbf{u})\right)^*\left(\left(\partial_p\mathcal{P}(S(\mathbf{u}),\mathbf{u})\right)^*\right)^{-1}.   
\end{equation}
Moreover, let $f$ be the solution of the following adjoint problem:
\begin{equation}
    \left(\partial_p\mathcal{P}(S(\mathbf{u}),\mathbf{u})\right)^* f=-N'(S(\mathbf{u})).
\label{adjointderiv}
\end{equation}
Combining the expressions \eqref{eq:Su_star} and \eqref{adjointderiv} with \eqref{eq:GradHatCF}, the gradient of the reduced cost function can be simplified to:
\begin{equation}
    \nabla \hat{J}=\left(\nabla_{\mathbf{u}}\mathcal{P}(S(\mathbf{u}),\mathbf{u})\right)^* f .
    \label{gradientderiv}
\end{equation}
By taking into consideration  $\displaystyle \Omega=\prod_{i=1}^n (0,a_i)$, with $a_i>0$ the adjoint problem \eqref{adjointderiv} can be explicitly given as:
\begin{equation}
    \mathcal{A}(f,\mathbf{u})=\begin{cases}
    \begin{array}{ccl}
       -\dfrac{\partial f}{\partial t}(t,\mathbf{x})&=& \displaystyle-\Gamma \cdot \nabla  f + \sum_{i=1}^{n}\left(k_m^i c_i(\mathbf{x};\mathbf{u})\int_{x_i}^{a_i} \! \omega_i(y_i-x_i)f(t,\mathbf{y}_i) \, \mathrm{d}y_i\right) \\
         && \displaystyle-\sum_{i=1}^{n} k_m^i  \!
         c_i(\mathbf{x},\mathbf{u})f(t,\mathbf{x}), 
    \end{array}  \\
	f(t_f,\mathbf{x})=-\left(p(t_{f}, \cdot )-p_{d}(\cdot )\right), \; \; \mathbf{x} \in \Omega,\\
   f(t,\mathbf{x})=0, \; \; (t,\mathbf{x}) \in \Sigma, 
    \end{cases}
    \label{adjoint}
\end{equation}
which evolves backward in time. Note the similarity with the formulation of  first passage times (see \cite{Pajaro:22}). The explicit formula of the gradient of the reduced cost function is obtained from \eqref{gradientderiv} as follows:
\begin{equation}
    \nabla \hat{J}(\mathbf{u})=-\int_{\Omega} \sum_{i=1}^{n}\left(k_m^i \int_0^{x_i} \! \beta_i(x_i-y_i)\nabla_{\mathbf{u}} \left[ c_i(\mathbf{y}_i;\mathbf{u})\right]p(t,\mathbf{y}_i) \, \mathrm{d}y_i\right)f(t,\mathbf{x})\mathrm{d}\mathbf{x} .
    \label{gradient}
\end{equation}
More details about the derivation of the adjoint problem from \eqref{adjointderiv} and the gradient of $\hat{J}$ are provided in Appendix S1.

Here it is important to remark that $\mathbf{u}$  can be any square integrable function in $U_{ad}$.   This includes the case of constant functions and piecewise constant functions which we will use next for Model Predictive Control. For these cases, we can directly derive the gradient by expression \eqref{gradient} by considering the following inner product,
\begin{equation} \label{eq:innerProdU}
 \left( \phi, f\right)_U=\int_{I}\int_{\Omega}\phi(t,\mathbf{x})f(t,\mathbf{x})\mathrm{d}\mathbf{x}\mathrm{d}t,   
\end{equation}
 where $U=L^2(I\times \Omega)$. Thus, the integration of the gradient, in both time and space, for the case of a constant input is given by:
 \begin{equation}
    \nabla \hat{J}(\mathbf{u})=-\int_{I} \int_{\Omega} \sum_{i=1}^{n}\left(k_m^i \int_0^{x_i} \! \beta_i(x_i-y_i)\nabla_{\mathbf{u}} \left[ c_i(\mathbf{y}_i;\mathbf{u})\right]p(t,\mathbf{y}_i) \, \mathrm{d}y_i\right)f(t,\mathbf{x})\mathrm{d}\mathbf{x}\mathrm{d}t.
    \label{constantgradient}
\end{equation}
For the case of piecewise constant function, where $\mathbf{u}$ is defined in $[t_0,t_f)$ with constant values in $[t_0,t_1),[t_1,t_2),...,[t_{n-1},t_f)$ the gradient on each time interval reads:
\begin{multline}
    \nabla \hat{J}(\mathbf{u})_{[t_k,t_{k+1})}= \\
    -\int_{t_k}^{t_{k+1}} \int_{\Omega} \sum_{i=1}^{n}\left(k_m^i \int_0^{x_i} \! \beta_i(x_i-y_i)\nabla_{\mathbf{u}} \left[ c_i(\mathbf{y}_i;\mathbf{u})\right]p(t,\mathbf{y}_i) \, \mathrm{d}y_i\right)f(t,\mathbf{x})\mathrm{d}\mathbf{x}\mathrm{d}t.
    \label{piecewiseconstantgradient}
\end{multline}

\subsubsection*{Discretization of the adjoint equation}

To solve the main initial value problem \eqref{maineq} we employ the  semi-Lagrangian numerical method proposed by \cite{Pajaro:17}. We  follow an optimize-before-discretize approach \cite{Borzi:12}. 
In order to avoid any inconsistency between the discretized objective and the reduced gradient we apply the same numerical scheme, spatial and temporal discretization to both the main equation \eqref{mainp} and the adjoint equation \eqref{adjoint}, with a finer mesh.

We consider a rectangular  computational domain $\Omega=\prod_1^n (0,a_i)$, with $a_i>0$. For the time discretization we consider a uniform mesh, where for $N>0$, the time step is $\Delta t:=\frac{t_f-t_0}{N}$, and the time mesh points are given by $t_q=t_0+q \Delta t$, for $q=0,1...,N$.  In order to derive a semi-Lagrangian scheme for the optimality system we introduce the material derivative of $p$ associated to the  vector field $-\Gamma$ as:
$$
\frac{D p}{D t}=\frac{\partial p}{\partial t}+\Gamma \cdot \nabla_{\mathbf{x}} p.
$$
This material derivative can be approximated in a point $(t_{q+1},\mathbf{x})$ as 
$$
\frac{D p}{D t}(t_{q+1},\mathbf{x}) \approx \frac{p(t_{q+1},\mathbf{x})-p(t_{q},\mathbf{x}_q)}{\Delta t},
$$
where $\mathbf{x}_q$ is the point occupied at time $t_q$ by a point placed in $\mathbf{x}$ at time $t_{q+1}$ that follows the characteristic trajectory defined by the vector field $-\Gamma$. Let $z(\mathbf{x},t_{q+1}; \cdot)$ be the characteristic curve, which is obtained as the solution to the following ordinary differential equation:
$$
\begin{cases}
    \dfrac{D z(\mathbf{x},t_{q+1}; \tau)}{D \tau}=-\Gamma(z(\mathbf{x},t_{q+1}; \tau)),\\
    z(\mathbf{x},t_{q+1}; t_{q+1} )=\mathbf{x}.
\end{cases}
$$
Thus, we have that $\mathbf{x_q}=z(\mathbf{x},t_{q+1}; t_{q})$. Using this notation, Pájaro et al \cite{Pajaro:17} provide the following formula to approximate the main equation \eqref{maineq} for every mesh point $\mathbf{x}$:
    \begin{equation}
        p(t_{q+1},\mathbf{x})=\frac{p(t_{q},\mathbf{x_q})+ \Delta t I(p)(t_{q},\mathbf{x})}{\displaystyle 1+\Delta t\left(\sum_{i=1}^{n} k_m^i  \! c_i(\mathbf{x})-\nabla_x \cdot \Gamma(\mathbf{x})\right)}.
    \end{equation}
Next, we follow the same methodology to obtain a semi-Lagrangian scheme to the adjoint equation \eqref{adjoint}.  As discussed before, it evolves backward in time. Therefore, we proceed similar to \cite{Pajaro:22}, transforming a backward problem into a forward one in order to simplify its approximation and calculation. We consider the time variable  change $t = t_f -s$, $s\in[t_0,t_f]$  such that $t\in I_F:=[0,t_f-t_0]$, $\phi(t,\mathbf{x})=f(s,\mathbf{x})$ and
$$\frac{\partial \phi}{\partial t}(t,\mathbf{x})=-\dfrac{\partial f}{\partial s}(s,\mathbf{x}).$$ 
Thus, we transform the backward adjoint equation \eqref{adjoint} into the following forward adjoint equation:  
\begin{equation}
    \mathcal{Y}(\phi,u)=\begin{cases}
    \begin{array}{cl}
       \dfrac{\partial \phi}{\partial t}(t,\mathbf{x})=& \displaystyle-\Gamma \cdot \nabla  \phi + \sum_{i=1}^{n}\left(k_m^i c_i(\mathbf{x},\mathbf{u})\int_{x_i}^{a_i} \! \omega_i(y_i-x_i)f(t,\mathbf{y}_i) \, \mathrm{d}y_i\right) \\
         & \displaystyle-\sum_{i=1}^{n} k_m^i  \!
         c_i(\mathbf{x},\mathbf{u})\phi(t,\mathbf{x}), 
    \end{array}  \\
	\phi(0,\mathbf{x})=-\left(p(t_{f}, \cdot )-p_{d}(\cdot )\right), \; \; \mathbf{x} \in \Omega,\\
   \phi(t,\mathbf{x})=0, \; \; (t,\mathbf{x}) \in \Sigma_F, 
    \end{cases}
    \label{adjointforward}
\end{equation}
where $\Sigma_F=I_F \times \partial^+ \Omega$. In this case the material derivative is associated to the vector field $\Gamma$, and the characteristic curves are described by the following ODE 
$$
\begin{cases}
    \dfrac{D z(\mathbf{x},t_{q+1}; \tau)}{D \tau}=\Gamma(z(\mathbf{x},t_{q+1}; \tau))\\
    z(\mathbf{x},t_{q+1}; t_{q+1} )=\mathbf{x}.
\end{cases}
$$
With the same argument as for the main equation \eqref{IBP} we obtain the following semi-Lagrangian scheme for the adjoint equation.
\begin{proposition}
    For every mesh point $\mathbf{x}$, we have the following approximation formula of the main equation \eqref{maineq}
    $$
    \phi(t_{q+1},\mathbf{x})=\frac{\phi(t_{q},\mathbf{x}_q)+ \Delta t \hat{I}(\phi)(t_{q},\mathbf{x})}{\displaystyle 1+\Delta t\sum_{i=1}^{n} k_m^i  \! c_i(\mathbf{x})}
    $$
    where 
    $$
    \hat{I}(\phi)=\sum_{i=1}^{n}\left(k_m^i c_i(\mathbf{x};\mathbf{u})\int_{x_i}^{a_i} \! \omega_i(y_i-x_i)\phi(t,\mathbf{y}_i) \, \mathrm{d}y_i\right).
    $$
\end{proposition}
To complete the numerical scheme, we discretize the protein space $\Omega=\prod_1^n (0,a_i)$, with a constant mesh step $\Delta \mathbf{x}=(\Delta x_1,...,\Delta x_n)$ for $\Delta x_i=a_i/N_i$ for $N_i\in \mathbb{N}$, $i=1,\dots,n$ and $N_i+1$ being the number of discrete protein points. The developed semi-Lagrangian scheme, along with the corresponding characteristic equations, is then used to find the values at each point $x_i$.

\subsubsection*{A nonlinear conjugate gradient scheme}
Due to the nonlinear nature of the optimization problem \eqref{OC}, we use in this section a Nonlinear Conjugate Gradient (NCG) optimization method which is an extension of linear conjugate gradient methods to nonquadratic problems (see for example \cite{Borzi:12}).  For $\mathbf{g}_{k+1}=\nabla_{\mathbf{u}} \hat{J}\left(\mathbf{u}_{k+1}\right)$, $\mathbf{g}_k \in \mathbb{R}^m, \ k=0,1,2,\dots,m$ the search direction, $\mathbf{d}_k \in \mathbb{R}^m, \ k=0,1,2,\dots,m$ is updated with the following formula:
\begin{equation} \label{eq:SearchDir}
    \mathbf{d}_{k+1}=-\mathbf{g}_{k+1}+\beta_k \mathbf{d}_k \; \; \; \;\mathbf{d}_{0}=-\mathbf{g}_{0},
\end{equation}
where $\beta_k$ is the conjugate gradient parameter. Different formulas for this parameter are proposed in the literature (see for example \cite{Hager:06}), here we use the formula introduced by Hager and Zhang in \cite{Hager:05}:
\begin{equation}
    \begin{cases}
    \mathbf{y}_k=\mathbf{g}_{k+1}-\mathbf{g}_{k},\\
    \beta_k=\frac{1}{\mathbf{d}_k^{\top} \mathbf{y}_k}\left(\mathbf{y}_k-2 \mathbf{d}_k \frac{\left\|\mathbf{y}_k\right\|^2}{\mathbf{d}_k^{\top} \mathbf{y}_k}\right)^{\top} \mathbf{g}_{k+1}.
\end{cases}
\label{pr}
\end{equation}
Through our simulations, this formula demonstrated superior effectiveness compared to other formulations for our specific case. Consequently, the optimization variable is updated with the formula
$$
\mathbf{u}_{k+1}=\mathrm{Proj}_{U_{ad}}(\mathbf{u}_k+\alpha_k \mathbf{d}_k),
$$
where the projection function is defined as 

$$
\mathrm{Proj}_{U_{ad}}(u_j)=\max(u^a_j,\min(u_j,u^b_j)).
$$ 

The positive step size $\alpha_k=\arg\min_\alpha \hat{J}\left(\mathbf{u}_k+\alpha \mathbf{d}_k\right)$ is obtained through a line search algorithm. We use the backtracking line search method, described in Algorithm \ref{Backtrack}, which is based on the Armijo–Goldstein condition \cite{Borzi:12}:
\begin{equation}\label{eq:AGcond}
    \hat{J}(\mathbf{u}_k+\alpha_k \mathbf{d}_k)\leq \hat{J}(\mathbf{u}_k)+\delta \alpha_k \left(\nabla_{\mathbf{u}} \hat{J}\left(\mathbf{u}_k\right), \mathbf{d}_k\right)_U \ \text{ for } \ \delta \in (0,\frac{1}{2}),
\end{equation}
with $(\cdot,\cdot)_U$ being the inner product defined in \eqref{eq:innerProdU}. 

The methodology consists of the  three algorithms bellow. Algorithm \ref{alg:NCG} encompasses the main Nonlinear Conjugate Gradient (NCG) iterations, beggining with an initial guess $\mathbf{u}_0$ for the optimal control variable and the computation of the initial gradient. Then the algorithm iterates through the NCG steps as described by expressions \eqref{eq:SearchDir}, \eqref{pr} and \eqref{eq:AGcond}. The stopping criteria include reaching the maximal number of iterations $N_{max}$,  a cost function value $\hat{J}\left(\mathbf{u}_k\right)$ at step $k$ is less than or equal to $tol_1$, or if the norm of the gradient is sufficiently small ($\leq tol_2$), indicating proximity to a minimum.

\begin{algorithm}[H]
\caption{Nonlinear Conjugate Gradient}\label{alg:NCG}
\begin{algorithmic}
  \State Input: initial approx. $\mathbf{u}_0, \mathbf{d}_0=-\nabla_{\mathbf{u}} \hat{J}\left(\mathbf{u}_0\right)$,\\
  Maximum number of iterations: $N_{\max }$\\
  Tolerance: $tol_1$ and $tol_2$
\While{$(k<N_{\max }$ \&\& $\hat{J}\left(\mathbf{u}_k\right)> tol_1 $ \&\& $\|\nabla_{\mathbf{u}} \hat{J}\left(\mathbf{u}_{k}\right)\|\mpd{_2}>tol_2$)}
  \State Perform a line search: Algorithm \ref{Backtrack} to solve $\alpha_k=\arg\min_\alpha \hat{J}\left(\mathbf{u}_k+\alpha \mathbf{d}_k\right)$
  \State Set $\mathbf{u}_{k+1}=\mathrm{Proj}_{U_{ad}}(\mathbf{u}_k+\alpha \mathbf{d}_k)$
   \State Evaluate $\nabla_{\mathbf{u}} \hat{J}\left(\mathbf{u}_{k+1}\right)$ with Algorithm \ref{alg:Gradeval} 
   \State Compute $\beta_k$ with formula \eqref{pr}
  \State Let $\mathbf{d}_{k+1}=-\nabla_{\mathbf{u}} \hat{J}\left(\mathbf{u}_{k+1}\right)+\beta_k \mathbf{d}_k$
   \State Set $k=k+1$

\EndWhile
\end{algorithmic}
\end{algorithm}

\begin{algorithm}[H]
\caption{Backtracking search for the Armijo–Goldstein condition}
\label{alg:Armijo}
\begin{algorithmic}
 \State Input: $\alpha_1=\min \left(\alpha_0, \left\|\mathbf{u}_k-\mathbf{u}_{k-1}\right\| /\left\|\mathbf{d}_k\right\|\right)$ and $\bar{\alpha}=0$, $\mathbf{u}_k$, $\mathbf{d}_k$, $i_{max}$, $tol_u$, $\delta$ 
 \While{$\left\|\alpha_i \mathbf{d}_k\right\|>\left\|\mathbf{u}_k\right\| tol_{\mathbf{u}} $  or $i<i_{max} $}
  \State Evaluate $\hat{J}\left(\mathbf{u}_k+\alpha_i \mathbf{d}_k\right)$
  \State if $\hat{J}\left(\mathbf{u}_k+\alpha_i \mathbf{d}_k\right) \leq \hat{J}\left(\mathbf{u}_k\right)+\delta \alpha_i\left(\nabla \hat{J}\left(\mathbf{u}_k\right), \mathbf{d}_k\right)_U$ set $\bar{\alpha}=\alpha_i$ break 
  \State else $\alpha_{i+1} \leftarrow \alpha_i / 2$
   \State if $(\bar{\alpha}==0)$ set error: step not found
\EndWhile
\end{algorithmic}
\label{Backtrack}
\end{algorithm}

\begin{algorithm}[H]
\caption{Gradient evaluation at $\mathbf{u}_k$}
\label{alg:Gradeval}
\begin{algorithmic}
  \State Solve the main IBP \eqref{mainp} with input $\mathbf{u}_k$ to obtain $p$
  \State Solve the adjoint equation \eqref{adjoint} with input $\mathbf{u}_k$ final condition $p(t_f,.)-p_d(.)$
  \State Use equation \eqref{gradient} to compute $\nabla_{\mathbf{u}} \hat{J}\left(\mathbf{u}\right)$ 
\end{algorithmic}
\end{algorithm}

\subsection*{Model Predictive Control with sampling window}
In this section we develop a  Model Predictive Control scheme (MPC) for stochastic gene regulatory networks.

\begin{figure}[H]
    \centering
   \includegraphics[scale=0.75]{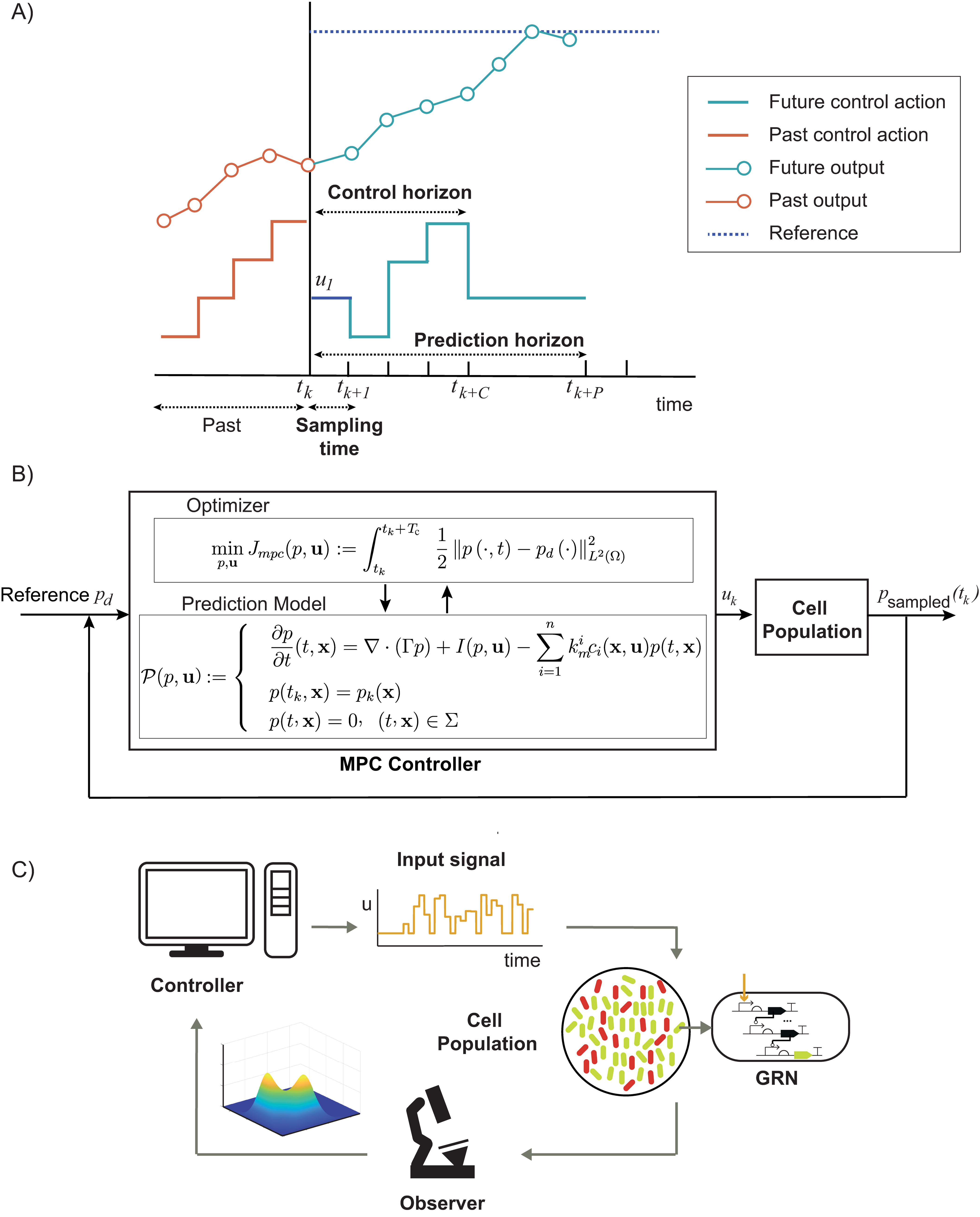}
    \caption{Schematic representation of the proposed MPC of gene regulatory networks  A) General scheme of MPC described in Algorithm \ref{alg:NMPC}; B) Scheme of MPC of gene regulatory networks via probability distribution function; C) Scheme of the proposed control realization configuration.}
    \label{fig:mpc}
\end{figure}

In practice, this is achieved by first defining a sampling time period $T$. Starting from an initial time point $t_0$, the subsequent sampling time points at which the state of the controlled system is measured are given by $t_k=t_0+kT$ for $k=1,\dots$. The core concept of MPC is to use a mathematical model to predict the future behavior of the controlled system over a time horizon $T_p$ at each sampling instant. At each sampling time point, an optimal control problem  is solved over a control horizon $T_c$.  The first value of the obtained optimal input profile is applied until the next sampling time point enabling effective feedback control. This process is described in Figure \ref{fig:mpc}.

To develop a MPC scheme we redefine the optimal control to be solved at each sampling step $t_k$ over a control horizon  $T_c=n_c T$, where $n_c$ is the number of sampling periods $T$. This ensures that  the difference between the solution of the prediction model \eqref{IVP} and the reference probability distribution is minimized at each sampling time step, rather than only at the final time. We also consider a piecewise control input $\mathbf{u}=\{\mathbf{u}_j\}_{j=1}^{n_c}$  where $\mathbf{u}_j$ are the constant values of $\mathbf{u}$ in the interval $[t_{k+j-1}, t_{k+j}[$, for $j=1,\dots, In_c$. In this case the optimal control problem is given by:
\begin{align}
	\min_{p,\mathbf{u}} J_{mpc}(p, \mathbf{u})&:=\int_{t_k}^{t_{k+n_c}}\frac{1}{2}\left\|p\left(\cdot, t\right)-p_{d}\left(\cdot\right)\right\|_{L^{2}(\Omega)}^{2} ,\label{mainmpOC}
 \end{align}
 \begin{align}
	 \mathcal{P}(p,\mathbf{u})&:=  \begin{cases}
	&\displaystyle \dfrac{\partial p}{\partial t}(t,\mathbf{x})  =  \nabla \cdot (\Gamma p)+I(p,\mathbf{u})- \sum_{i=1}^{n} k_m^i  \! c_i(\mathbf{x},\mathbf{u})p(t,\mathbf{x}), \\
	&p(t_k,\mathbf{x})=p_k(\mathbf{x}), \\
 &p(t,\mathbf{x})=0, \; \; (t,\mathbf{x}) \in \Sigma .
 \end{cases}\label{mainmpc}
\end{align}

Following the same arguments as in the previous section, we obtain the following adjoint problem:
\begin{equation}
    \mathcal{A}_{mpc}(f,\mathbf{u})=\begin{cases}
    \begin{array}{cl}
         -\dfrac{\partial f}{\partial t}(t,\mathbf{x})  =& \displaystyle -\Gamma \cdot \nabla  f + \sum_{i=1}^{n}\left(k_m^i c_i(\mathbf{x},\mathbf{u})\int_{x_i}^{a_i} \! \omega_i(y_i-x_i)f(t,\mathbf{y}_i) \, \mathrm{d}y_i\right)\\
         & \displaystyle - \sum_{i=1}^{n} k_m^i  \! c_i(\mathbf{x},\mathbf{u})f(t,\mathbf{x})
         -\left(p(t, \mathbf{x} )-p_{d}(\mathbf{x})\right),
    \end{array}\\
	f(t_{k+n_c} ,\mathbf{x})=-\left(p\left(t_{k+n_c}, \cdot \right)-p_{d}\left(\cdot \right)\right) \; \; \mathbf{x} \in \Omega,\\
   f(t,\mathbf{x}) =0, \; \; (t,\mathbf{x}) \in \Sigma . \\
    \end{cases}
    \label{adjointmpc}
\end{equation}
The gradient of the cost function on each variable control input $\mathbf{u}_j$   corresponding to each time interval $[t_{k+j-1}, t_{k+j}[$ is given by:
\begin{multline}
    \nabla \hat{J}(\mathbf{u})_{[t_{k+j-1}, t_{k+j}[}=\\
    -\int_{t_{k+j-1}}^{t_{k+j}} \int_{\Omega} \sum_{i=1}^{n}\left(k_m^i \int_0^{x_i} \! \beta_i(x_i-y_i)\nabla_u \left[ c_i(\mathbf{y}_i,\mathbf{u})\right]p(t,\mathbf{y}_i) \, \mathrm{d}y_i\right)f(t,\mathbf{x})\mathrm{d}\mathbf{x}\mathrm{d}t
    \label{gradientmpc}
\end{multline}
Note that this definition also allows for the possibility of defining multiple target distributions at various time points by replacing $p_d(.)$ with a time-variant target distribution $p_d(t,.)$.  The MPC related optimal control problem \eqref{mainmpOC} can be solved following Algorithm \ref{alg:NCG} described in the optimal control section with the new adjoint \eqref{adjointmpc} and gradient \eqref{gradientmpc} formulations. This process is then iterated through the sampling times  $t_k,\;k=0,...,N_{max}$ as described in Algorithm \ref{alg:NMPC}.
\begin{algorithm}
\caption{Nonlinear Model Predictive Control Algorithm}
\label{alg:NMPC}
\begin{algorithmic}
  \State Let $t_k$ for $k=0,1,...$ be the sampling time points 
  \State Set $n_c T$ the control horizon
  \State Set $p(t_0,x)=p_0$ the initial state
  \State Set the target distribution $p_d(.)$
  \For{$k=0,1,...N_{max}$}
  \State Measure the state $p(t_k,\mathbf{x})$ of the system
  \State Use Algorithm \ref{alg:NCG} to solve OCP in $(t_k,t_{k+n_c})$  to obtain  the optimal input  $\Bar{\mathbf{u}}$ 
  \State Use the first time point $u=\Bar{\mathbf{u}}(t_k)$ till the next sampling period
  \EndFor
\end{algorithmic}
\end{algorithm}

\subsection*{Computational Reachability}
Before applying Algorithms \ref{alg:NCG} to \ref{alg:NMPC}, we need to test whether given the initial state  $P_0$ there exists a control input driving the system to the target distribution $P(t_f)$ at time $t_f$. Here we propose a computational approach to test whether the target state is reachable at time $t_f$, as outlined in Algorithm \ref{alg:reachability}. 

First, we provide the Reachability Algorithm \ref{alg:reachability} with a target distribution $p_d$ as input and consider a fixed initial time $t_0$ (usually taken as 0) along with initial final time $T_f^0$. Using  Algorithms \ref{alg:NCG} to \ref{alg:Gradeval} we solve the optimal control problem \eqref{mainp} with $T_f^0$ as the final time. We then update the final time by adding a time step $\Delta t$.  This process is repeated until either the maximum final time $T_f^{max}$ is reached or the norm $\left\|p\left(\cdot, T_f^k\right)-p_{d}\left(\cdot\right)\right\|_{L^{2}(\Omega)}^{2}$ falls below a  predefined tolerance $tol$. The parameter $tol$ represents the degree of closeness to the original target distribution. If this tolerance is not achieved, we select the state with the minimal cost function, denoted as $J^{opt}$ as defined in the algorithm. This closest reachable distribution at time $t_f$ is  then used as the target for further application of the the MPC algorithm. 

Note that, for this reachability test, the only constraints on the input profile are that it must be square-integrable and within the bounds $(\mathbf{u}^L,\mathbf{u}^U)$. This extends the range of control input profiles used for the test. In the next Section we will illustrate the performance of the approaches and algorithms we have developed through two case studies. All computations were performed on a MacBook Pro with Apple M1 Max chip with 10 cores and 3.22GHz, and 64 GB of RAM, using MATLAB R2022a running under MacOS Sonoma 14.4.1 version.

\begin{algorithm}[H]
\caption{Reachability }\label{alg:reachability}
\begin{algorithmic}
  \State Input:\\
  target distribution $p_d$\\
  Initial and maximal final time  $T_f^{0}$ and $T_f^{\max }$\\
  Time step: $\Delta t$\\
  Tolerance: $tol$ 
\While{$(T^k_f<T_f^{\max }$ \&\& $\hat{J}^{opt}> tol )$}
   \State Set $T^{k+1}_f=T^k_f+\delta t$;
   \State Apply Algorithm \ref{alg:NCG} to solve the OCP for $T_f^{k+1}$ to obtain $\left(\hat{J}_{T^{k+1}_f},\mathbf{u}_{T^{k+1}_f}\right)$ 
   \If{$\hat{J}_{T^{k+1}_f}<\hat{J}_{T^{k}_f}$}
   \State $\hat{J}^{opt}=\hat{J}_{T^{k+1}_f}$
   \EndIf
\EndWhile
\end{algorithmic}
\end{algorithm}

\section*{Results and Discussion}

\subsection*{Achieving bimodality in bacterial populations through MPC}

In our first example, the goal is to induce bimodality in a population of isogenic cells endowed with a an self-regulatory circuit motif that is not inherently bimodal. Positive autoregulation is one of the simplest motifs capable of demonstrating bimodality, allowing the cell to stochastically switch between two states. Consequently, controlling such motifs has a wide range of applications. While the positive autoregulatory structure can generate bistable behavior, this property is highly dependent on the kinetic parameter values \cite{Pajaro:15}. The dynamics of a stochastic gene positive autoregulatory network repressed by an external signal are described by the PIDE model \eqref{maineq}:
\begin{align} 
\frac{\partial p}{\partial t}(t, \mathbf{x})=&  \left(\frac{\partial}{\partial x}\left[\gamma_x x p(x)\right]\right.  \left.+k_m \int_0^{x} \omega\left(x-y\right) c\left(y,u\right) p\left(t, y\right) \mathrm{d} y-k_m c(x,u) p(x)\right).
\label{pide_autoregulation}
\end{align}
The input function $c(x,u)$ is given by
\begin{equation*}
c(x,u)=\epsilon(1-\rho(x,u))+ \rho(x,u)
\end{equation*}
with:
\begin{equation*}
\rho(x,u)=\frac{\left(\frac{x}{K}h(u)\right)^{n_h}}{1+\left(\frac{x}{K}h(u)\right)^{n_h}}
\end{equation*}
and 
\begin{equation*}
    h(u)=\left(\frac{1}{1+\frac{u}{K_u}}\right)^{n_u}
\end{equation*}
and $\omega$ in \eqref{pide_autoregulation} reads:
$$
\omega_i\left(x-y\right)=\frac{1}{b} \exp \left(-\frac{x-y}{b}\right)\; \; \; \;\textnormal{and}\; \; \; \;b=\frac{k_x}{\gamma_m},
$$
the parameters $k_x,\gamma_x,\;k_m\; and\; \gamma_m $ \eqref{paramters_autoregulation} are the production and degradation rates of $mRNA$ and its corresponding protein respectively. 

\begin{figure}[H]
\includegraphics[width=\linewidth]{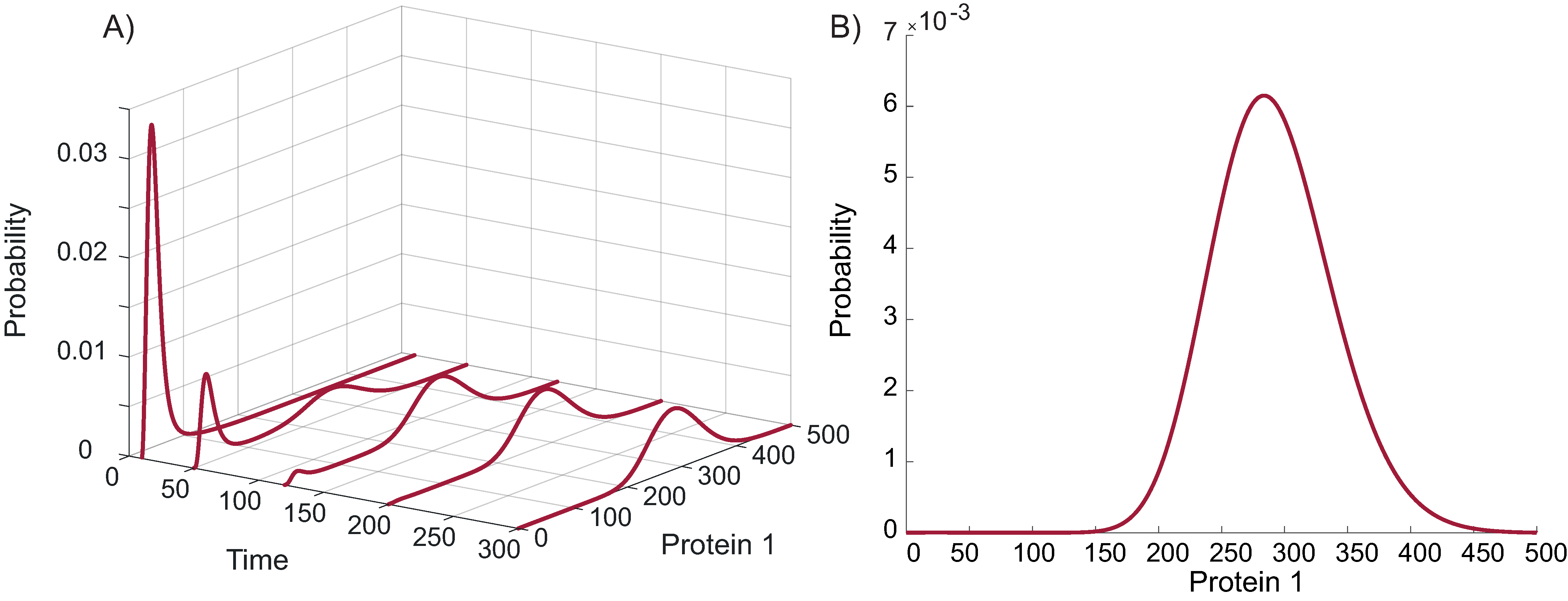}
\caption{Solution of the PIDE model \eqref{pide_autoregulation} in absence of control: A) snapshots at different time points, showing transient bimodality that disappears before reaching the stationary state; B) steady state of the PIDE model at $t=300$ min.}
\label{fig:unimodal}
\end{figure}

Taken with parameter values listed in Table \ref{paramters_autoregulation},  the stationary distribution of the gene regulatory network is unimodal, as illustrated in Figure \ref{fig:unimodal} which depicts the dynamics and stationary distribution of the uncontrolled system. Our goal is driving the cell population to  a bi-modal state using an external signal.

\begin{table}[ht]
\begin{adjustwidth}{-2.25in}{0in} 
\centering
\caption{
{\bf Parameter values for the  positive autoregulation network}}
\begin{tabular}{|l|l|l|}
\hline
\multicolumn{1}{|l|}{\bf Parameter} & \multicolumn{1}{|l|}{\bf Description} & \multicolumn{1}{|l|}{\bf Value}\\ \thickhline
$k_m $& Transcription kinetic constant &20  \\ 
 \hline
 $\gamma_m$& mRNA degradation constant & 0.5  \\
 \hline
 $k_x$& Translation kinetic constant & 2.5  \\
 \hline
 $\gamma_x$ & Protein degradation constant & 0.35  \\
 \hline
 $K$ & Hill kinetic coefficient & 80  \\
 \hline
 $K_u$ & Hill kinetic coefficient (signal) &200  \\
 \hline
 $n_h$ & Cooperativity &4  \\
 \hline
  $n_u$ & Cooperativity (signal) &3  \\
 \hline 
 $\epsilon$ & Transcription leakage & 0.1 \\ \hline
\end{tabular}
\begin{flushleft} 
\end{flushleft}
\label{paramters_autoregulation}
\end{adjustwidth}
\end{table}

\subsubsection*{Reachability of the  bi-modal state} 

\begin{figure}[H]
\includegraphics[scale=0.75]{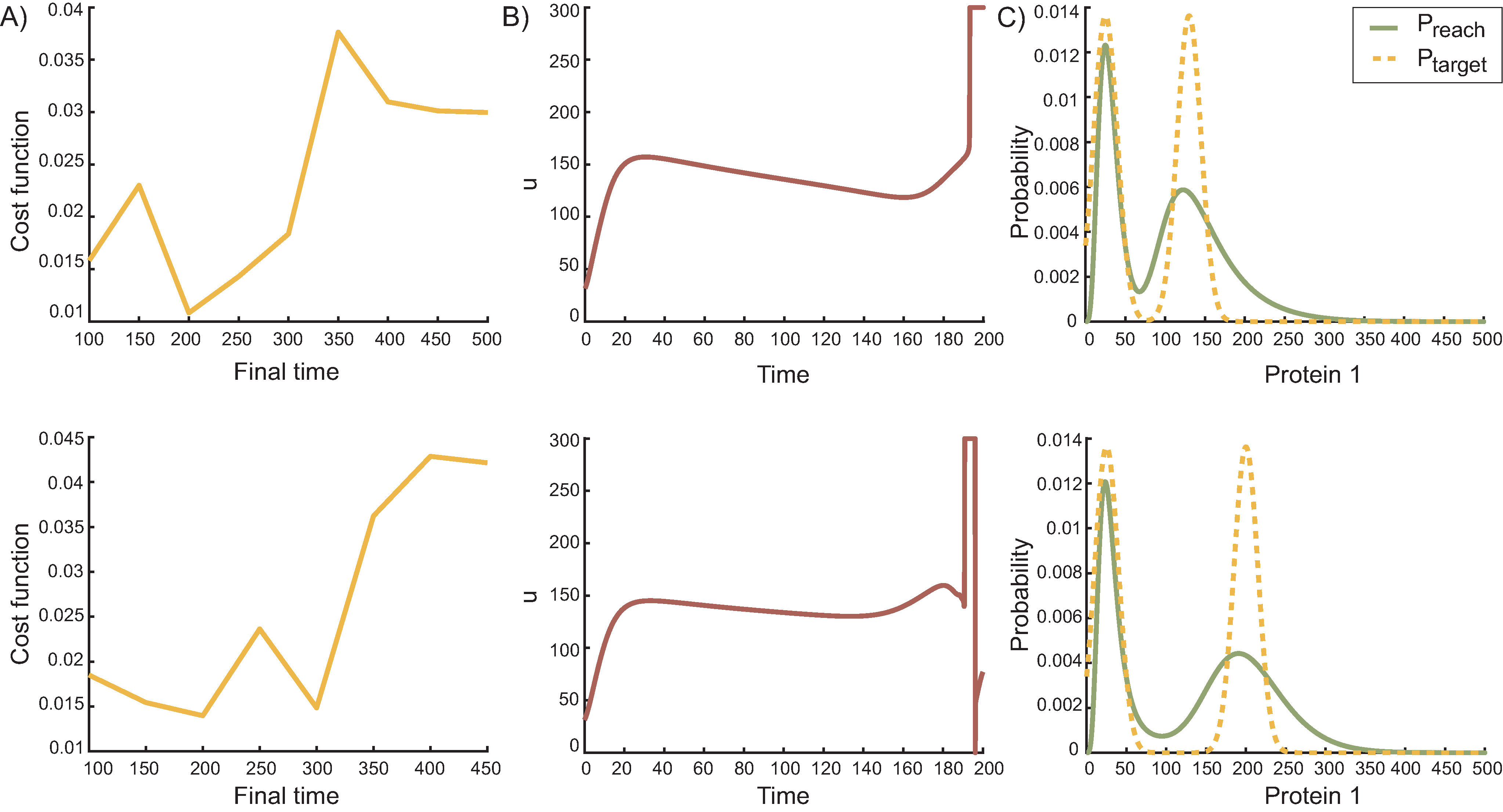}
\caption{Reachability results for two different target distributions. 
The results for the first target distribution (mode 2 = 130, $\gamma=15$) are shown in the top row, while the results for the second target distribution (mode 2 = 200, $\gamma=15$) are shown in the bottom row: column A) optimal values of the cost function  for each final time $T_f$, the cost reaches its minimum at $T_f = 200$; column B) optimal input profile obtained ($T_f = 200$); column C) closest reachable solution to the target distribution  $(T_f = 200)$. }
\label{fig:reachability}
\end{figure}

First, based on Algorithm \ref{alg:reachability}, we set a target bimodal state $p_d$ depicted by the dashed lines in Figure \ref{fig:reachability}. By applying the algorithm,  we obtain the closest reachable state $P_{reach}$,  shown by the green solid lines, which we will further use as reference target. The choice of $p_d$ is based on its modes; throughout out this example, the first mode at $x=25$ remains fixed  whereas the second mode is varied to test reachability.

In most cases tested, we obtain a $P_{reach}$ with the same modes as $p_d$, but with different variability around the modes.  This demonstrates the  effectiveness of the optimal control method. We illustrate the results for two different target bimodal distributions in Figure  \ref{fig:reachability}.

The computational performance of the reachability test was evaluated under various conditions, specifically considering the second mode $(M_2)$ and the variability around this mode, denoted by $\gamma$, in the time interval $T_f^0=100$ and $T_f^{max}=450 ~min$ as described in Algorithm \ref{alg:reachability}. For the pairs ($M_2=130$, $\gamma=25$), ($M_2=130$, $\gamma=15$), ($M_2=200$, $\gamma=15$), and ($M_2=250$, $\gamma=15$). The computation times were consistent across the tests, with an average completion time of approximately 3.300 seconds.

\subsubsection*{Bi-modality through Model Predictive Control}
Next, we apply the MPC Algorithm \ref{alg:NMPC} to reach the reference distribution $P_{reach}$ with with two modes at $x_1=25$ and value $x_2=200$ at final time $t_f$, and stabilise the network around it. 

\begin{figure}[H]
\centering
\includegraphics[scale=0.75]{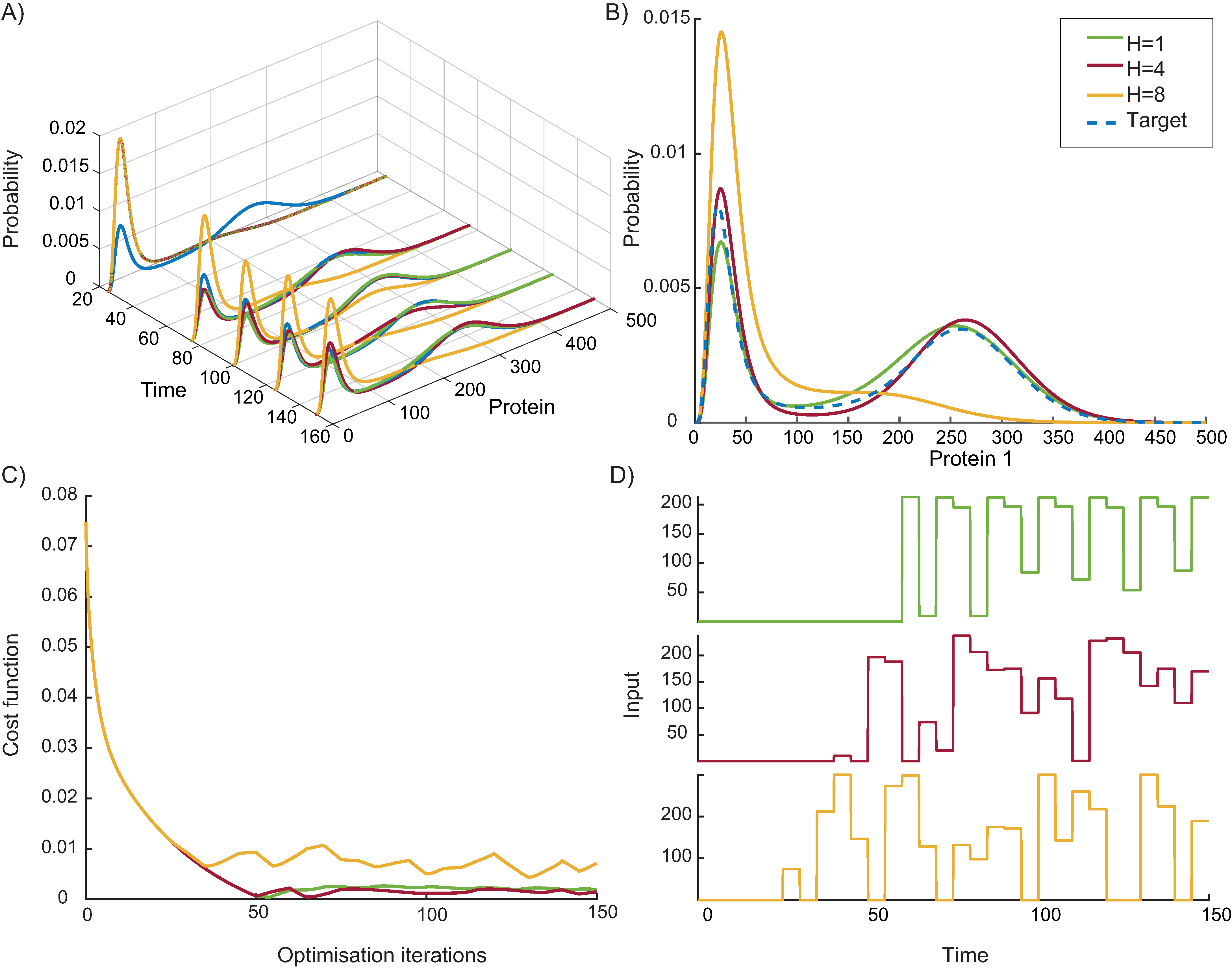}
\caption{Performance of MPC method for $P_{reach}$  with two modes at $x_1=25$ and value $x_2=250$  under different prediction horizons: A) evolution of the solutions for different sampling times under control; B) solution of the PIDE model at $t=150$ min for different $H$ with $P_{reach}$ in dashed lines; C) cost function $\max_{x \in \Omega}|p(t,.)-p_d(.)|$ used to evaluate the performance at each sampling period; D) input profile for different $H$. }
\label{fig:mpc_horizon}
\end{figure}

In Figure \ref{fig:mpc_horizon} we show the results obtained for different control horizons ($H=1, H=4, H=8$) with a sampling time $T=5$. Figures \ref{fig:mpc_horizon} A) and B) show respectively the dynamic evolution of the different solutions and the solution at $t_f=150$ min, to evaluate the performance of the controller we use the we use the max error value $\max_{x \in \Omega}|p(t,.)-p_d(.)|$ at each time point t, this value is plot in Figure C) of each panel and referred to as cost function. Clearly, short and medium term horizons $H=1$ and $H=4$ show better performances. This counterintuitive observation is most likely due to the highly stochastic nature of the system, the computational performance for each control horizon is given in table \ref{fig:mpc_time_onegene}.


The MPC algorithm demonstrates high computational efficiency, producing the results shown in Fig. \ref{fig:mpc_horizon} in 104 seconds for $H=1$, 440 seconds for $H=4$ and 1073 seconds for $H=8$.

\subsection*{Tracking moving target probability distributions in bacterial populations}
 
In this section we apply the previously discussed control algorithms to the a gene inducible network based on \cite{Milias-Argeitis:16} for the control of gene expression and cell growth. Our ultimate goal is for the bacterial population to be able to track moving target distribution. The dynamics of the inducible network is described by the PIDE model \eqref{maineq}, with the input function given by:
$$
c(x,u)=\epsilon+\frac{K_u}{k_m}u,
$$
where u is the external signal acting as an input to the network. The parameters are provided in Table \ref{fig:opto_parameters}.

\begin{table}[ht]
\begin{adjustwidth}{-2.25in}{0in} 
\centering
\caption{
{\bf Kinetic parameters for the positive autoregulation network (time units in minutes)}}
\begin{tabular}{|l|l|}
\hline
\multicolumn{1}{|l|}{\bf Parameters} & \multicolumn{1}{|l|}{\bf Values}\\ \thickhline
$k_m $& 0.0048  \\ 
 \hline
 $\gamma_m$ & 0.0048  \\
 \hline
 $k_x$& 0.0116  \\
 \hline
 $\gamma_x$ & 0.0016  \\
 \hline
 $K_u$ & 0.0965  \\
 \hline
 $\epsilon$ & 0.5  \\ 
 \hline
\end{tabular}

\label{fig:opto_parameters}
\end{adjustwidth}
\end{table}

\begin{figure}[H]
\includegraphics[scale=0.75]{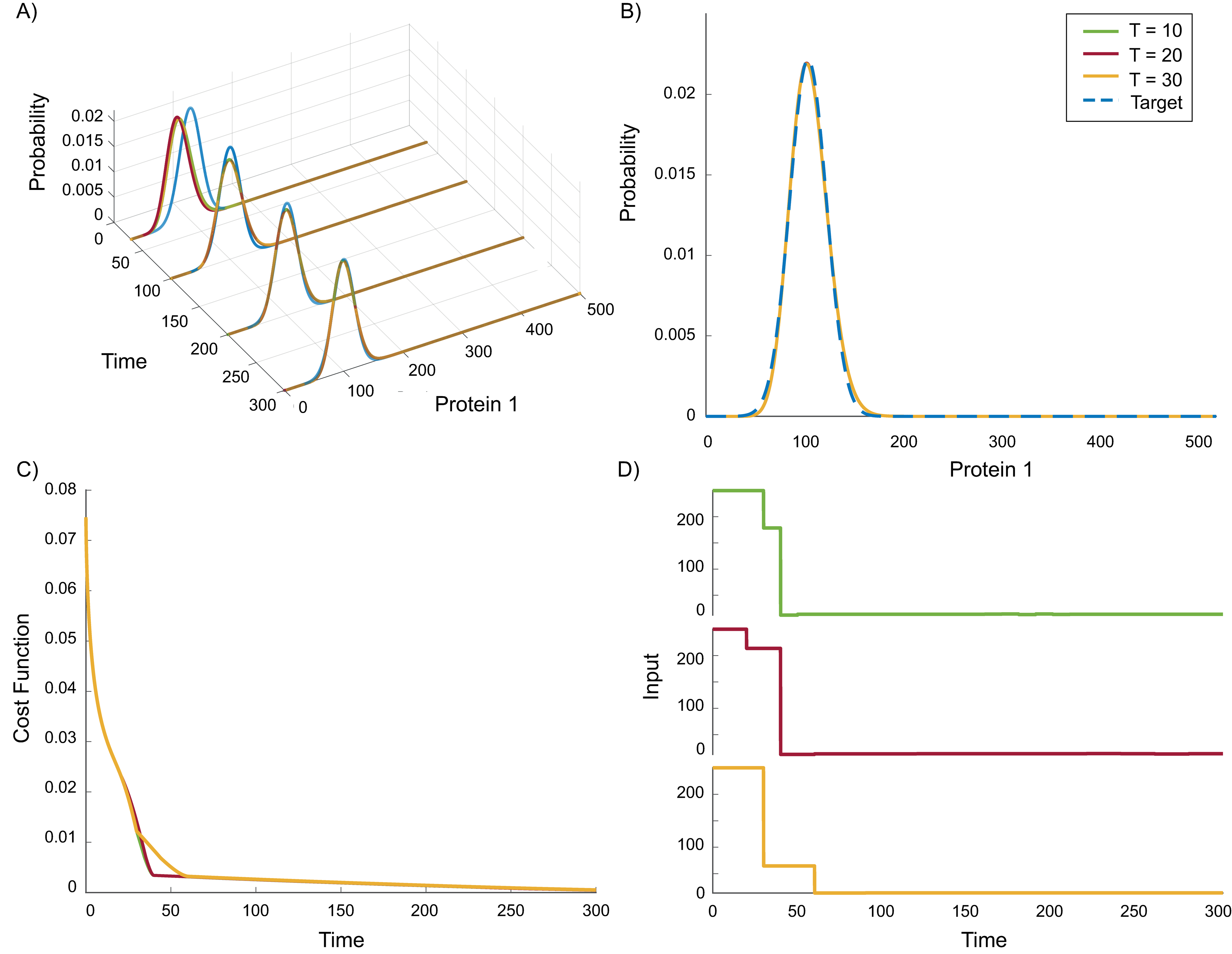}
  \caption{Results for different sampling times ranging from 10 to 30 min: (A) Temporal evolution of the PIDE solution for different sampling intervals; (B) Mode simulations showing the target distribution at $t_f=300$ min for each sampling time; C) cost function $\max_{x \in \Omega}|p(t,.)-p_d(.)|$ to evaluate performance  for different sampling times; D) input profiles corresponding to the different sampling times.}
  \label{fig:opto_mpc_results_sample}
\end{figure}

First, in a similar way as the previous example, we perform a reachability test to obtain reachable distributions. Next, we solve the simpler problem of stabilise the network around a given $P_{reach}$ by applying Algorithm \ref{alg:NMPC}. To evaluate the performance of the method we compare the effect of several prediction horizons H  and sampling times T. In Figure \ref{fig:opto_mpc_results_sample} we depict the results obtained for different sampling times, showing a similar trend to the previous example. The main difference lies in the time required to initially reach the target: longer sampling times result in a longer period to reach the target. However, once the target is reached, there are no significant differences in performance.  Further results about effect of the  prediction horizon and computational performance are given in Appendix S1. 

The MPC algorithm produced results showin in Fig. \ref{fig:opto_mpc_results_sample} (with $T_f=300$) for sampling times 10, 20 and 30 in approximately 100 seconds.

\subsubsection*{Tracking several distributions}
\begin{figure}[H]
\includegraphics[scale=0.75]{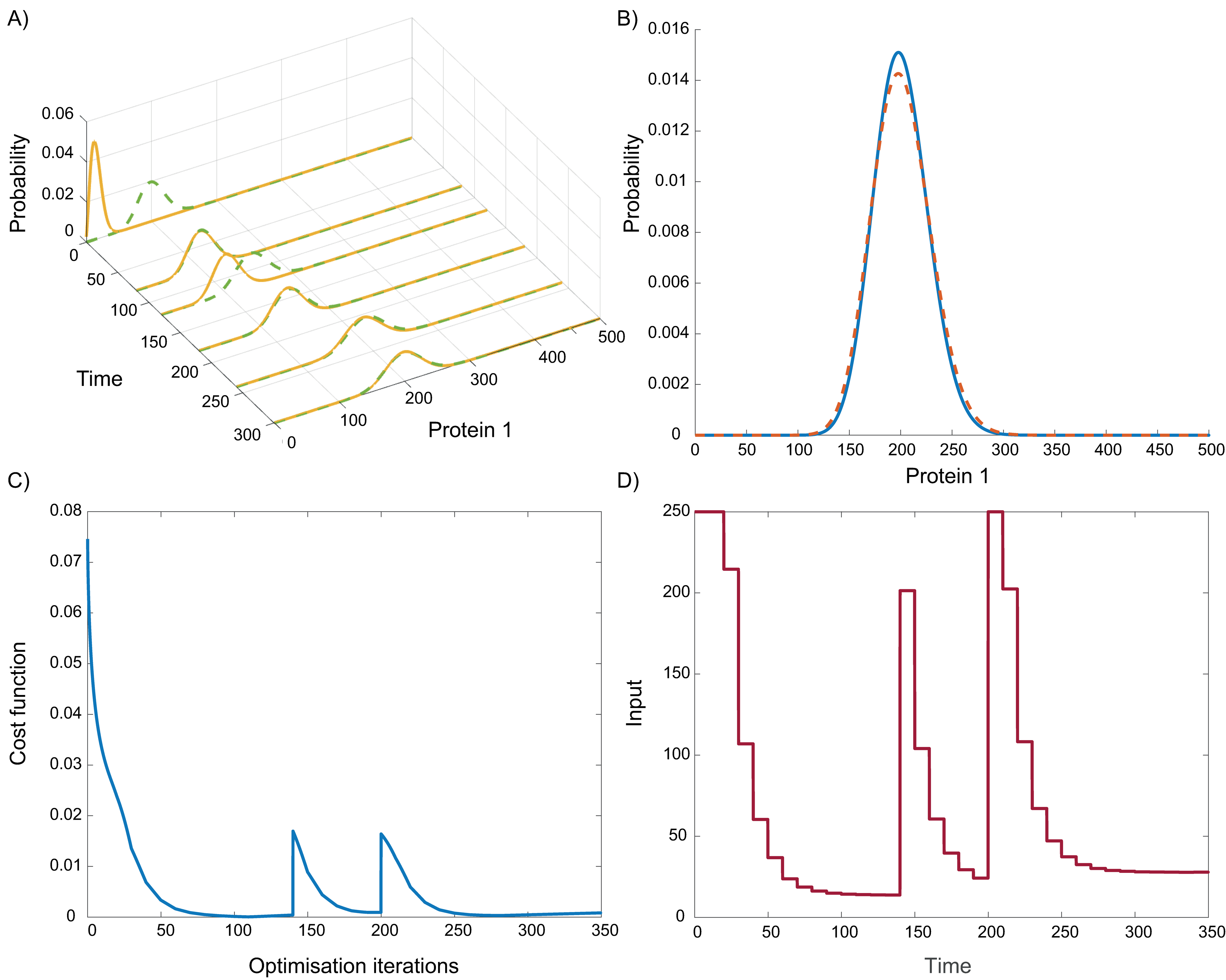}
  \caption{Results of tracking moving distributions A) Evolution of the PIDE solutions as the input value is updated to track the new target; B) Final result at time 350 for the last target state C) cost function $\max_{x \in \Omega}|p(t,.)-p_d(.)|$ D) input profile.}
      \label{fig:opto_results_track}
\end{figure}

We now address the challenge of tracking multiple references at different time points. Specifically, we aim to track three different distributions at predefined intervals, as depicted in Figure \ref{fig:opto_results_track}. The target is modified at $t_1=150$ min and at $t_2=200$ min. The algorithm effectively tracks the changes in $P_{reach}$ and, as observed in Figure \ref{fig:opto_results_track} (A, B) the new targets are almost perfectly achieved. This demonstrates the robustness of the developed MPC algorithm in tracking varying targets. We have employed a sampling time $T=10$ min and a prediction horizon of $H=20$ min. In Figure  \ref{fig:opto_results_track} C peaks in the cost function are evident: when the target is changed, the cost function is reset to the new reference resulting in the observed increases.

\section*{Conclusion}

In this work, we establish a general framework for optimal control and Model Predictive Control of gene regulatory networks. Using Partial Integro-Differential Equations (PIDE) have significant advantages for controlling stochastic systems, especially those where the molecular noise is high (as it is usually the case in bacteria and yeast). 

One of the key advantages of PIDE based control is its ability to transform the control of a purely stochastic system into the control of a deterministic model. By focusing on the entire probability distribution over time, our PIDE based control framework not only manages the moments of the distribution but also allows for the manipulation of more complex shapes and behaviors. Unlike the  Fokker Plank approximation associated to the Chemical Langevin Equation, the PIDE approach is particularly well-suited for handling stochastic systems showing multimodality and other nonlinear emergent properties. The validation of the PIDE approximation against the Gillespie algorithm has been demonstrated in previous studies
\cite{Pajaro:17}, confirming its accuracy and reliability.

Furthermore, the explicit formulation of the gradient of the cost function at every discretization point presents a powerful method for solving large-scale optimal control problems. This adjoint-based approach proves more efficient than standard optimization toolboxes, particularly when dealing with the complexities of reachability analysis. The computational methods employed here are highly efficient and effectively address the challenges posed by theoretical complexities.

It is also important to note the flexibility of our approach, where the control function $\mathbf{u}$  can be any square-integrable function within the admissible set in $U_{ad}$ \eqref{eq:AdmiSp}. This flexibility enhances the applicability of our methods to a wide range of control problems in stochastic systems.

Using the MPC approach we developed, we demonstrate the ability to drive a gene regulatory network toward bimodality. By adjusting the control input  while keeping the kinetic parameters fixed, we can reach different bimodal states. Bimodality plays a crucial role in cellular decision-making under molecular noise, and the ability to engineer and regulate these responses through control mechanisms may have significant implications in biology.

\section*{Supporting information}

\paragraph*{S1 Appendix.}
\label{S1_Appendix}
Proofs and additional properties of GRN-PIDEs.

\paragraph*{S1 Code.}
\label{S1_Code}
Matlab code to reproduce the results in the paper.

\section*{Acknowledgments}
HF acknowledges funding from e-MUSE MSCA-ITN-2020 European Training Network under the Marie Sklodowska-Curie grant agreement No 956126. IOM acknowledges grant PID2021-127888NA-I00 (COMPSYNBIO) funded by MCIN/AEI/10.13039/501100011033, grant TED2021-131049B-
I00 (BioEcoDBTL) funded by MCIN/AEI/10.13039/501100011033 and NextGenerationEU/PRTR
and grant CIAICO/2021/159 (SmartBioFab) funded by Generalitat Valenciana.
MP acknowledges funding by the Spanish Ministry of Science and Innovation through project PID2022-141058OB-I00, as well as from the Galician Government through grant ED431C 2022/47, both of which include FEDER financial support.

\nolinenumbers

%
%
%

\newpage
{\Large\textbf\newline{S1 Appendix: Supporting Information }}

\renewcommand{\thesection}{S1.\arabic{section}} 
\renewcommand{\theequation}{S1.\arabic{equation}}
\renewcommand{\thetheorem}{S1.\arabic{theorem}}
\section{Some properties of the GRN PIDE solution}
In this section we include some fundamental properties of the solution of the GRN PIDE model  \eqref{maineq} and the Initial Value Problem \eqref{IVP}. The well-posedness of \eqref{IVP} is proven in \cite{SIbib3}  in the generalized (mild) sense, that is, for $p_0 \in \mathcal{L}^1\left(\mathbb{R}^n\right)$ there exists a unique mild solution $p \in \mathcal{C}\left(\mathbb{R}_{+} ; L^1\left(\mathbb{R}^n\right)\right)$. Moreover, the following lemma is verified \cite{SIbib3}:
\begin{lemma}
Given any mild solution $p$  of \eqref{IVP}, we have:\\
(i) If $p_0(\mathbf{x})\geq 0, \; \; \forall \mathbf{x} \in \mathbb{R}^n $, then  $p(t,\mathbf{x})\geq 0 \; \; \forall t \geq 0$ with p being the solution of the initial value problem defined in \eqref{IVP}.\\
(ii) Mass conservation:
$$
\int_{\mathbb{R}_{+}^n} p(t, \boldsymbol{x}) \mathrm{d} \boldsymbol{x}=\int_{\mathbb{R}_{+}^n} p_0(\boldsymbol{x}) \mathrm{d} \boldsymbol{x}.
$$\\
(iii) The $L^1$ norm is decreasing
$$
\left\|p(t)\right\|_1 \leq\left\|p_0\right\|_1, \quad \forall t \geq 0.
$$
The existence of a classical solution is proven  if $p_0 \in \mathcal{C}^{1, b}\left(\mathbb{R}_{+}^n\right)$   then there exists a unique classical solution $p \in \mathcal{C}^1\left(\mathbb{R}_{+} ; L^1\left(\mathbb{R}_{+}^n\right)\right)$ with $b>0$. 
\end{lemma}

We include the control $\mathbf{u}$ in the definition of the initial boundary problem given in the main text, expression \eqref{IBP}:
\begin{equation}
  \mathcal{P}(p)=  \begin{cases}
	&\dfrac{\partial p}{\partial t}(t,\mathbf{x})  =  \nabla \cdot (\Gamma p)+I(p,\mathbf{u}) - \displaystyle \sum_{i=1}^{n} k_m^i  \! c_i(\mathbf{x},\mathbf{u})p(t,\mathbf{x}), \\
	&p(t_0,\mathbf{x})=p_0(\mathbf{x}), \\
 &p(t,\mathbf{x})=0, \; \; (t,\mathbf{x}) \in \Sigma,
 \end{cases}
 \label{SI_IBP}
\end{equation}
with $\mathbf{u}$ defined as in the main text in $$U_{ad}=\{\mathbf{u} \in [L^2(I)]^m : u_j^a \leq u_j \leq u_j^b, \ j=1,\dots,m \} \subset [L^2(I)]^m$$
 where $\mathbf{u}^a,\mathbf{u}^b\in \mathbb{R}^m$ and $u^a_j \leq u^b_j, \ j=1,\dots,m$ are the control constrains.\\
 Next, we develop some a priori estimates that will be useful to prove the well posedness of the formulated optimal control problem. We  Consider the following functional spaces  $H:=L^2(\Omega)$.  Where for bounded Lipschitz domain V is defined as:
 \begin{align*}
     V &=\{ f \in H^1(\Omega) : f(x)=0 \text{ on } \partial^+ \Omega \},\\
     H^1(\Omega)&=\{ f \in L^2(\Omega): \partial_{x_i} f \in L^2(\Omega) \text{ for } i=1,....n \}.
 \end{align*}
 with 
 $$
 \langle f,g\rangle_V := \int_{\Omega} f \cdot g \mathrm{d}\mathbf{x}+ \int_{\Omega} \nabla f \cdot  \nabla g \mathrm{d}\mathbf{x}\
 $$
 
 For more details about these functional spaces we refer the reader to \cite{SIbib4}. $V^*$ denotes the dual space of $V$  defined as the space of all continuous linear mappings from $V$ to $\mathbb{R}$ we consider the duality pairing  $\langle f, g\rangle_{V^* V}:=  \int_{\Omega} fg \,\mathrm{d}\mathbf{x}$ for $f \in V^*$ and $g \in V$. We also define:
$$
\begin{aligned}
L^2(I ; V)&=\left\{f: I \rightarrow V \text { such that } \int_I\|f(t)\|_V^2 \mathrm{d} t<\infty\right\}
\end{aligned}
$$
which is a Banach space, equipped with the following norm:
$$
\|f\|_{L^2(I ; V)}:=\left(\int_I\|f(t)\|_V^2 \mathrm{~d} t\right)^{\frac{1}{2}}, \text{with } ||.||_V:= \left(\int_{\Omega}|\nabla f(t,.)|^2 \mathrm{d}\mathbf{x} \right)^{\frac{1}{2}}.
$$
We define the following Hilbert space
\begin{equation} 
    F=\left\{f \in L^2(I ; V) \text{ such that } \partial_t f \in L^2(I ; V^*)\right\},
\end{equation}
which is equipped with the  inner product 
$$
\langle f, g \rangle_{F}=\int_I \langle f,g\rangle_V \mathrm{d}t+\int_I \langle \partial_t f, \partial_t g \rangle_{V^*} \mathrm{d}t,
$$
For the following proposition we omit the dependence of c on $\mathbf{u}$ and c is written simply as $c_i(x)$.
\begin{proposition}
\label{estimates}
Let $p_0 \in V$, assuming that $c_i$, $i=1,\dots,n$ are positive continuous bounded functions, there exist positive constants $M_i>0$ such that: 
$$
\forall \mathbf{x} \in \Omega,\; \; c_i(\mathbf{x}) \leq M_i, \ \forall i.
$$
So that, if $p$ is a solution to \eqref{IVP}, then there exist $K_1>0$ and $K_2>0$, such that the following inequalities hold
\begin{align}
    &\|p\|_{L^2(I ; V)}\leq K_1 \|p_0\|_{L^2(\Omega)}, \nonumber\\
    &\|\partial_t p\|_{L^2(I ; V^*)}\leq K_2\|p_0\|_{L^2(\Omega)}.
\end{align}

\begin{proof}
We start proving the proposition for the one dimensional case. 
By multiplying the equation \eqref{maineq} by $p$ and integrating each term over $\Omega$ we obtain:
\begin{dmath}
   \int_{\Omega} \partial_t p(t,x) p(t,x) \mathrm{d}x+\int_{\Omega}k^1_m c_1(x)p^2(t,x)\mathrm{d}x=\int_{\Omega} \partial_x (\gamma^1_x x p(t,x))p(t,x)\mathrm{d}x+\int_{\Omega}I(p)p(t,x)\mathrm{d}x.
    \label{weakformu_aux}
\end{dmath}
We introduce the pairing $\langle . , .\rangle_{V^* V}$ to denote the first term in \eqref{weakformu_aux}, which can be rewritten as:
\begin{equation}
    \langle \partial_t p , p\rangle_{V^* V}+\int_{\Omega}k^1_mc_1(x)p^2(t,x)\mathrm{d}x=\int_{\Omega} \partial_x (\gamma^1_x x p(t,x))p(t,x)\mathrm{d}x+\int_{\Omega}I(p)p(t,x)\mathrm{d}x.
    \label{weakformu}
\end{equation}
First, as proven in \cite{SIbib4}, we have that:
\begin{equation} \label{eq:partialTerm}
    \langle \partial_t p , p\rangle_{V^* V}=\dfrac{1}{2}\dfrac{d}{dt}\|p(t,.)\|^2_{L^2(\Omega)}.
\end{equation}
Moreover, the second term in \ref{weakformu} is positive and:
\begin{equation} \label{eq:positiveTerm}
    \int_{\Omega}k^1_mc_1(x)p^2(t,x)\mathrm{d}x \leq k^1_m M_1 \|p\|^2_{L^2(\Omega)}.
\end{equation}
Next, we deal with each term of the equality \ref{weakformu} separately. Firstly, we notice that 
\begin{align*}
\int_{\Omega} \gamma^1_x x \partial_x (p(t,x))p(t,x)\mathrm{d}x
&=\frac{ \gamma^1_x}{2}\int_{\Omega} x \partial_x (p(t,x)^2) \mathrm{d}x,
\end{align*}
and integrating by parts we obtain: 
\begin{align*} 
\int_{\Omega} \gamma^1_x x \partial_x (p(t,x))p(t,x)\mathrm{d}x=\frac{ \gamma^1_x}{2} \left( \left[x p(t,.)^2\right]_{\Sigma}-\int_{\Omega} (p(t,x))^2 \mathrm{d}x\right)
&=-\frac{ \gamma^1_x}{2}\int_{\Omega} (p(t,x))^2 \mathrm{d}x.
\end{align*}
Using the last equality we obtain the following result
\begin{align*}
\int_{\Omega} \partial_x (\gamma^1_x x p(t,x))p(t,x)\mathrm{d}x
&=\gamma^1_x \int_{\Omega} (p(t,x))^2 \mathrm{d}x + \int_{\Omega} \gamma^1_x x \partial_x (p(t,x))p(t,x) \mathrm{d}x\\
&=\gamma^1_x \int_{\Omega} (p(t,x))^2 \mathrm{d}x-\frac{ \gamma^1_x}{2}\int_{\Omega} (p(t,x))^2 \mathrm{d}x\\
&=\frac{ \gamma^1_x}{2} \int_{\Omega} (p(t,x))^2 \mathrm{d}x.
\end{align*}
Therefore, we have that:
\begin{equation} \label{eq:degradationTerm}
\left|\int_{\Omega} \partial_x (\gamma^1_x x p(t,x))p(t,x) \mathrm{d}x\right| \leq \frac{ \gamma^1_x}{2}  \|p\|^2_{L^2(\Omega)}.
\end{equation}
Now, we consider the integral term in \eqref{weakformu}: 
\begin{align*}
    \int_{\Omega}I(p)p(t,x) \mathrm{d}x 
    &=\int_{\Omega} k_m^1 \int_0^{x} \! \omega_1(x-y)c_1(y)p(t,y) \, \mathrm{d}y\, p(t,x) \mathrm{d}x\\
    &=\int_{\Omega} \frac{k_m^1}{b_1} \int_0^{x} \! e^{\frac{y}{b_1}}c_1(y)p(t,y) \, \mathrm{d}y \, e^{-\frac{x}{b_1}} p(t,x) \mathrm{d}x.  
\end{align*}
By using the positivity property of $c$ and $p$, we obtain:
$$
 \int_0^{x} e^{\frac{y}{b_1}}c_1(y)p(t,y) \, \mathrm{d}y< \int_{\Omega} e^{\frac{y}{b_1}}c_1(y)p(t,y) \, \mathrm{d}y
$$

As the term on the right is constant (it does not depends on $x$) we obtain:
$$
\int_{\Omega}I(p)p(t,x) \mathrm{d}x \leq \frac{k_m^1}{b_1} \left(\int_{\Omega}e^{-\frac{x}{b_1}}p(t,x) \mathrm{d}x\right) \left(\int_{\Omega}e^{\frac{y}{b_1}}c_1(y)p(t,y) \mathrm{d}y\right).
$$
This leads to:
\begin{equation}\label{eq:integralTerm}
 \int_{\Omega}I(p)p(t,x)\mathrm{d}x \leq C_1 |\Omega|^2 \|p(t,.)\|^2_{L^2(\Omega)},   
\end{equation}
where $C_1= \dfrac{k_m^1}{b_1} e^{\frac{|\Omega|}{b_1}} M_1 $.

Combining the results \eqref{eq:partialTerm}, \eqref{eq:positiveTerm}, \eqref{eq:degradationTerm} and \eqref{eq:integralTerm} with the formulation \eqref{weakformu} we obtain that:
$$
\frac{d}{dt}\|p(t,.)\|^2_{L^2(\Omega)} \leq \frac{d}{dt}\|p(t,.)\|^2_{L^2(\Omega)} + k_m^1M_1  \|p(t,.)\|^2_{L^2(\Omega)} \leq K^1  \|p(t,.)\|^2_{L^2(\Omega)},
$$
with $K^1= \gamma_x^1+2C_1|\Omega|^2$.

Application of Gronwall inequality leads to first result:
\begin{equation}
    \|p\|_{L^2(I ; V)}\leq K_1^1 \|p_0\|_{L^2(\Omega)}.
    \label{estimate1}
\end{equation}
Following arguments given in \cite{SIbib4} we obtain the second inequality in the proposition. 

The result can be generalized to higher dimensions by summing the inequality terms and constants $K^i_1$ and $K^i_2$.
\end{proof}
\end{proposition}

\section{Some proofs and first order optimality conditions}
\renewcommand{\thetheorem}{\arabic{theorem}}
\setcounter{theorem}{0}

In this section we include the proofs for Proposition  \ref{diffcost} and for Theorem \ref{th:solUOP} from the main text. Let us start by Proposition  \ref{diffcost}:

\begin{proposition}
\label{SI_diffcost}
The control-to-state operator $\textit{S}:U_{ad}\rightarrow F $ is Fréchet differentiable, and the directional derivative in the direction $\mathbf{h}\in[L^2(I)]^m$ is given by
$$
\nabla_{\mathbf{u}} S(\mathbf{u})\cdot \mathbf{h}=g,
$$
where $g$ is  the weak solution to the initial-boundary value problem:
\begin{align} \label{SI_Fr}
&  \begin{array}{ll}
      \displaystyle \dfrac{\partial g}{\partial t}(t,\mathbf{x})  = &  \displaystyle \nabla \cdot (\Gamma g)+ I(g,\mathbf{u})-\sum_{i=1}^{n} k_m^i  \!c_i(\mathbf{x},\mathbf{u})g(t,\mathbf{x})+\nabla_{\mathbf{u}} I(p,\mathbf{u})\cdot \mathbf{h}(t)\\
       &  \displaystyle -\sum_{i=1}^{n} k_m^i  \! \nabla_{\mathbf{u}} c_i(\mathbf{x},\mathbf{u})\cdot \mathbf{h}(t)p(t,\mathbf{x}),  \nonumber
   \end{array} \\
	&g(t_0,\mathbf{x})=0, \\
 &g(t,\mathbf{x})=0, \; \; (t,\mathbf{x}) \in \Sigma. \nonumber
\end{align}
\end{proposition}

\begin{proof}
  For the proof we will focus on one dimensional case. Following the methodology used in \cite{SIbib4}. Let $\mathbf{u},\mathbf{h} \in [L^{2}(I)]^m $, we denote by  $p$ the solution associated to $\mathbf{u}$, and by $p_{\mathbf{h}}$ the solution associated to $\mathbf{u+h}$, hence $p=S(\mathbf{u})$ and $p_{\mathbf{h}}=S(\mathbf{u}+\mathbf{h})$. Therefore, \eqref{SI_IBP}  is verified by $p$ and $p_{\mathbf{h}}$ for $\mathbf{u}$ and $\mathbf{u+h}$, respectively. By subtracting them, it yields:
  \begin{equation} \label{eq:prop1_PIDE}
     \begin{array}{cl}
     \displaystyle \partial_t (p_{\mathbf{h}}-p)=   & \displaystyle \nabla \cdot (\Gamma (p_{\mathbf{h}}-p))+ I(p_{\mathbf{h}},\mathbf{u}+\mathbf{h})-I(p,\mathbf{u}) \\
         & \displaystyle - k_m^1 \left[  c_1(x,\mathbf{u}+\mathbf{h})p_{\mathbf{h}}(t,x)- c_1(x,\mathbf{u})p(t,x)\right].
    \end{array} 
  \end{equation}
   Next, we focus on the last term of the previous expression. By assumption the operator $c_1(x,.)$ is Fréchet differentiable, and we have that:
   $$
   c_1(x,\mathbf{u+h})=c_1(x,\mathbf{u})+\nabla_{\mathbf{u}} c_1(x,\mathbf{u})\cdot \mathbf{h}+r_1(\mathbf{h}),
   $$
   with a remainder $r_1$ such that:
   \begin{equation}\label{eq:remainder1}
     \frac{\|r_1\|_2}{\|\mathbf{h}\|_2} \rightarrow 0 \; \; \;\text{as} \; \; \;\|\mathbf{h}\|_2 \rightarrow 0.  
   \end{equation}
   Consequently, we obtain that:
   $$
   k_m^1 \left[ c_1(x,\mathbf{u+h})p_h- c_1(x,\mathbf{u})p\right]=k_m^1\left[c_1(x,\mathbf{u})(p_{\mathbf{h}}-p)+\nabla_{\mathbf{u}} c_1(x,\mathbf{u})\cdot \mathbf{h} p_{\mathbf{h}}+r(\mathbf{u},\mathbf{h})p_{\mathbf{h}} \right].
   $$
   Similarly, for the integral term $I(p,u)$ we obtain: 
   $$I(p_{\mathbf{h}},\mathbf{u}+\mathbf{h})-I(p,\mathbf{u})=I(p_{\mathbf{h}}-p,\mathbf{u})+\nabla_{\mathbf{u}}I(p_{\mathbf{h}},\mathbf{u})\cdot\mathbf{h}+\Tilde{I}(p_{\mathbf{h}},r_2(\mathbf{h})),$$
   where
   \begin{align*}
\nabla_{\mathbf{u}}I(p_{\mathbf{h}},\mathbf{u})\mathbf{h}&=k^1_m \int_0^{a_1}w(x-y)\nabla_{\mathbf{u}}c_1(x,\mathbf{u})\cdot \mathbf{h}(t) p_{\mathbf{h}}\mathrm{d}x, \\
\Tilde{I}(p_{\mathbf{h}},r_2(\mathbf{h})) &=k^1_m r_2(\mathbf{h}) \int_0^{a_1}w(x-y) p_{\mathbf{h}}\mathrm{d}x,
   \end{align*}
   and the remainder $r_2$ such that:
   \begin{equation}\label{eq:remainder2}
     \frac{\|r_2\|_2}{\|\mathbf{h}\|_2} \rightarrow 0 \; \; \;\text{as} \; \; \;\|\mathbf{h}\|_2 \rightarrow 0.  
   \end{equation}
Let $\Tilde{p}=p_{\mathbf{h}}-p$, then, by substituting the terms in the right hand side of expression \eqref{eq:prop1_PIDE} for the above approximations, one can easily check that $\Tilde{p}$  verifies the following PIDE: 
$$
   \begin{cases} 
   & \begin{array}{rl}
       \dfrac{\partial \Tilde{p}}{\partial t}(t,x)  = &  \nabla \cdot (\Gamma \Tilde{p})+ I(\Tilde{p},\mathbf{u})- k_m^1  \!c_1(x,\mathbf{u})\Tilde{p}(t,x)+\nabla_{\mathbf{u}} I(p_{\mathbf{h}},\mathbf{u})\cdot \mathbf{h}(t)\\
        & - k_m^1  \! \nabla_{\mathbf{u}} c_1(x,\mathbf{u})\cdot \mathbf{h}(t)p_{\mathbf{h}}(t,x)-r_1(\mathbf{h})p_{\mathbf{h}}+\Tilde{I}(p_{\mathbf{h}},r_2(\mathbf{h})),  \nonumber
   \end{array}   
   \\
	&\Tilde{p}(t_0,x)=0, \\
 &\Tilde{p}(t,x)=0, \; \; (t,x) \in \Sigma. \nonumber
\end{cases}
 $$  
   Let $w=\Tilde{p}-g$. By subtracting the expression \eqref{SI_Fr} (verified by $g$) from the last equation (verified by $\Tilde{p}$), we obtain that $w$ satisfies the next PIDE:
   $$
    \begin{cases} 
   & \begin{array}{rl}
       \displaystyle \dfrac{\partial w}{\partial t}(t,x)  = &  \displaystyle \nabla \cdot (\Gamma w)+ I(w,\mathbf{u})-k_m^1 \!c_1(x,\mathbf{u}) w (t,x)+\nabla_{\mathbf{u}} I(\Tilde{p},\mathbf{u})\cdot \mathbf{h}(t)\\
       & \displaystyle - k_m^1  \! \nabla_{\mathbf{u}} c_1(x,\mathbf{u})\cdot \mathbf{h}(t)\Tilde{p}(t,x)-r_1(\mathbf{h})p_{\mathbf{h}}+\Tilde{I}(p_{\mathbf{h}},r_2(\mathbf{h})),  \nonumber
   \end{array}  
   \\
	&w(t_0,x)=0, \\
 &w(t,x)=0, \; \; (t,x) \in \Sigma. \nonumber
\end{cases}
 $$   
Note that, this equation for $w$ is similar to the main problem \ref{SI_IBP} plus the term: 
$$\nabla_{\mathbf{u}} I(\Tilde{p},\mathbf{u})\cdot \mathbf{h}(t)
   - k_m^1  \! \nabla_{\mathbf{u}} c_1(x,\mathbf{u})\cdot \mathbf{h}(t)\Tilde{p}(t,x)-r_1(\mathbf{h})p_{\mathbf{h}}+\Tilde{I}(p_{\mathbf{h}},r_2(\mathbf{h})).$$
Then, by applying the estimate \eqref{estimate1} from the previous proposition \ref{estimates}, we have:
   $$
    \|w\|_{C^2(I ; H)}\leq K_1^1 \|w_0\|_{L^2(\Omega)}+R_1 \|\Tilde{p}\|_{L^2(I ; H)}^2 \|h\|_{L^2(I)}^2+(\|r_1\|_{L^2(I)}^2+R_2\|r_2\|_{L^2(I)}^2 )\|p_{\mathbf{h}}\|^2_{L^2(I ; H)},
   $$ 
with $R_1$ being a constant resulting 
 from the boundedness of  $\nabla_{\mathbf{u}} c_1$ (as in \ref{eq:integralTerm}) and from the following inequality 
$$
 \|\nabla_{\mathbf{u}} I(\Tilde{p},\mathbf{u})\cdot \mathbf{h}
   - k_m^1  \! \nabla_{\mathbf{u}} c_1(x,\mathbf{u})\cdot \mathbf{h}\Tilde{p} \|_{L^2(I ; H)}^2 \leq  \|\nabla_{\mathbf{u}} I(\Tilde{p},\mathbf{u})\cdot \mathbf{h}\|\ +  \|\ k_m^1  \! \nabla_{\mathbf{u}} c_1(x,\mathbf{u})\cdot \mathbf{h}(t)\Tilde{p}\|_{L^2(I ; H)}^2.
$$ 
Similarly the  constant $R_2$ is resulting from $\Tilde{I}(p_{\mathbf{h}},r_2(\mathbf{h}))$. 
Since $w_0=w(t_0,x)=0$ and assuming that $\|p_{\mathbf{h}}\|^2_{L^2(I ; H)}<\infty$, we have that: 
 $$
    \frac{\|w\|_{C^2(I ; H)}}{\|h\|_{L^2(I)}}\leq \|h\|_{L^2(I)}\left( R \|\Tilde{p}\|_{L^2(I ; H)}^2 +\left(\frac{\|r_1\|_{L^2(I)}^2}{\|h\|_{L^2(I)}^2}+\frac{\|r_2\|_{L^2(I)}^2}{\|h\|_{L^2(I)}^2} \right)\|p_{\mathbf{h}}\|_{L^2(I ; H)}^2\right)\rightarrow 0, 
   $$   
   as $\|h\|_{L^2(I)} \rightarrow 0$. Note that, the remainders $r_1$ and $r_2$ verify \eqref{eq:remainder1} and \eqref{eq:remainder2}, respectively.
\end{proof}

Next, we prove the theorem \ref{th:solUOP} of the main text:

\begin{theorem}
Let $p_0 \in V$, then there exist an optimal pair $(\Bar{p}, \Bar{\mathbf{u}})$ solution to the optimal control problem \ref{OC}, such that $\Bar{p}=S(\Bar{\mathbf{u}})$ is the unique solution to \ref{mainp} associated to $\Bar{\mathbf{u}}$,  and $\Bar{\mathbf{u}}$ is solution to the unconstrained optimization problem \ref{uncon}.
\end{theorem}
\begin{proof}
    By definition of the cost function,  the functional $\hat{J}$ is bounded from bellow, and so it has a finite infimum $j$. Therefore, there exist a minimizing sequence $\{\mathbf{u}_n\}_{n\in \mathbb{N}}$ such that $\displaystyle\lim_{n \to \infty}\hat{J}(\mathbf{u}_n)=j$. \\
    $U_{ad}$ is a convex, closed, and bounded subset of the reflexive Banach space $L^2(I)$. Thus, we can extract a subsequence $\{\mathbf{u}_k\}_{k\in \mathbb{N}}$ such that:
    $$
    \mathbf{u}_k \rightharpoonup \Bar{\mathbf{u}} \text{ as } k \rightarrow \infty,
    $$
    where $\Bar{\mathbf{u}} \in U_{ad}$ is admissible, due to the closeness and convexity of $U_{ad}$. The weakly convergent subsequence $\{\mathbf{u}_k\}_{k\in \mathbb{N}}$ generates a sequence $\{p_k\}_{k\in \mathbb{N}}$ of solutions to the initial boundary problem \eqref{SI_IBP} defined by $S(\mathbf{u}_k)=p_k\in F$, by virtue of Aubin-Lions lemma the embedding $F \hookrightarrow L^2(I,H)$ is compact. Therefore there exists a subsequence $\{p_i\}_{i\in \mathbb{N}}$ that converges strongly to $\Bar{p} \subset L^2(I;H)$.\\

   Thanks to the estimates in Proposition \ref{estimates}, the sequence $\partial_t p_s$ is bounded in $L^2(I;V)$, therefore it converges weakly to $\partial_t \Bar{p}$. Moreover, the smoothness of the integral operator $I: L^2(I;V)\times L^2(I) \to \mathbb{R}$ and function $c(\mathbf{x},.)$ ensure that  as $k,s \to \infty$ we have
    \begin{align*}
        I(p_s,\mathbf{u}_k) \to I(\Bar{p},\Bar{\mathbf{u}})\\
        c(\mathbf{x},\mathbf{u}_k) \to c(\mathbf{x},\Bar{\mathbf{u}}) 
    \end{align*}
The convergence of the initial and boundary conditions follows from standard arguments as discussed in \cite{SIbib4}. We conclude that $S(\Bar{\mathbf{u}})=\Bar{p}$ and the final step of the proof is to show that the pair $(\Bar{p},\Bar{\mathbf{u}})$ is a minimiser of the cost functional $J$. Due to the convexity of $J$, a direct application of Theorem 2.12 page 47 of \cite{SIbib4} ensures the  weakly lower semicontinuity of $J$, that is: 
$$
\lim_{s\to \infty} \inf J(p_s,\mathbf{u}_s)\geq J(\Bar{p},\Bar{\mathbf{u}}),
$$
which yields the result of the theorem.
\end{proof}

\subsection{First order optimality conditions}
In this section we give more details about the derivation of the adjoint problem \ref{adjoint} and the gradient formulation \eqref{gradient} given in the main text. The gradient of the cost function leads to the following equations (expressions \eqref{adjointderiv} and \eqref{gradientderiv} in the main text):
\begin{equation}
    \left(\partial_p\mathcal{P}(S(\mathbf{u}),\mathbf{u})\right)^* f=-N'(S(\mathbf{u})),
\label{SI_adjointderiv}
\end{equation}
where $f$ is the solution of the adjoint problem.  The gradient of the reduced cost function, is also given by:
\begin{equation}
    \nabla \hat{J}=\left(\nabla_{\mathbf{u}}\mathcal{P}(S(\mathbf{u}),\mathbf{u})\right)^* f .
    \label{SI_gradientderiv}
\end{equation}
By taking into consideration $\Omega=[0,a_1[\times[0,a_2[\dots\times[0,a_n[\subset \mathbb{R}^n$, 
the equation \eqref{SI_adjointderiv} can be explicitly formulated through the derivation of the adjoint of the operator $\partial_{p}\mathcal{P}(S(\mathbf{u}),\mathbf{u})$, for the simplicity of notation and clarity of the proof we consider one dimension. Let  $f \in F^*$  and $g \in F$, based on the definition of $\mathcal{P}(p,\mathbf{u})$ in \eqref{mainp}, we have: 
\begin{multline}
 <\partial_{p}\mathcal{P}(p,\mathbf{u})g,f>= \\ -\int_{t_0}^{t_f}\int_{0}^{a_1}\bigg[-\frac{\partial g}{\partial t}f  + \partial_{x} (\gamma_x^1 x g)f+I(g)f - \displaystyle k_m^1  \! c_1(x,\mathbf{u})g f \bigg]\mathrm{d}x\mathrm{d}t ,  
 \label{eq:SI_InnerP_fg}
\end{multline}
where each term of this equation can be treated separately. By integrating by parts with respect to $t$ the time derivative term in \eqref{eq:SI_InnerP_fg}, we obtain:
$$
\int_{t_0}^{t_f}\int_{0}^{a_1}-\frac{\partial g}{\partial t}f \mathrm{d}x \mathrm{d}t=\int_{t_0}^{t_f}\int_{0}^{a_1}\frac{\partial f}{\partial t}g \mathrm{d}x \mathrm{d}t  -\int_{0}^{a_1}\left[g(t_f,x)f(t_f,x)-g(t_0,x)f(t_0,x)\right]\mathrm{d}x.
$$
deriving the initial condition we have $g(t_0,x)=0$  therefore the  term $g(t_0,x)f(t_0,x)$ vanishes. For the second term integration by parts on $x$ leads to 
\begin{align*}
    \int_{t_0}^{t_f}\int_{0}^{a_1}\partial_{x} (\gamma_x^1 x g)f\mathrm{d}x\mathrm{d}t=&-\int_{t_0}^{t_f}\int_{0}^{a_1} \gamma_x^1 x \partial_{x}(f) g \mathrm{d}x\mathrm{d}t+\int_{t_0}^{t_f}\left[\gamma_x^1 x g(t,x)f(t,x)\right]_0^{a_1}\mathrm{d}t\\
    =&-\int_{t_0}^{t_f}\int_{0}^{a_1} \gamma_x^1 x\partial_{x} ( f)g \mathrm{d}x\mathrm{d}t,
\end{align*}
where the second equality is results of the boundary conditions as $g(t,a_1)=0$ and also for$x=0$ we have $\gamma_x^1 x g(t,x)f(t,x)=0$ (with the assumption that $|g(t,0)| |f(t,0)|<\infty$).
For the integral term we have
$$
\int_{t_0}^{t_f}\int_{0}^{a_1} I(g)f \mathrm{d}x\mathrm{d}t=\int_{t_0}^{t_f}\int_{0}^{a_1} \left(k_m^1\int_0^{x} \! \omega_1(x-y)c_1(y,\mathbf{u})g(t,y) \, \mathrm{d}y\right)f(t,x)\mathrm{d}x \mathrm{d}t.
$$
Note that, $0<y<x$ and $0<x<a_1$ so $y<x<a_1$. Thus, by applying Fubini's theorem to integrals on $x$ and $y$in the last expression, we obtain: 
\begin{align*}
    \int_{t_0}^{t_f}\int_{0}^{a_1} I(g)f \mathrm{d}x\mathrm{d}t&=\int_{t_0}^{t_f}\int_{0}^{a_1} \left(k_m^1\int_y^{a_1} \! \omega_1(x-y)c_1(y,\mathbf{u})f(t,x) \, \mathrm{d}x\right)g(t,y) \mathrm{d}y\mathrm{d}t\\
    &=\int_{t_0}^{t_f}\int_{0}^{a_1} \left(k_m^1 c_1(y,\mathbf{u})\int_y^{a_1} \! \omega_1(x-y)f(t,x) \, \mathrm{d}x\right)g(t,y) \mathrm{d}y\mathrm{d}t.
\end{align*}
We denote the new integral term by
$$
\hat{I}(f)(t,y)=k_m^1 c_1(y,\mathbf{u})\int_y^{a_1} \! \omega_1(x-y)f(t,x) \mathrm{d}x.
$$
Summing the previous expressions, which expand the terms on the right-hand side of expression \eqref{eq:SI_InnerP_fg}, we obtain:  
\begin{align*}
    <g,(\partial_{\mathbf{p}}\mathcal{P}(p,\mathbf{u}))^*f>=&\int_{t_0}^{t_f}\int_{0}^{a_1}\bigg[\frac{\partial f}{\partial t}-\gamma_x^1 x\partial_{x}  f+ \hat{I}(f)(t,x)- k_m^1  \! c_1(x;\mathbf{u})f\bigg]g(t,x) \mathrm{d}x\mathrm{d}t\\
&+\int_{0}^{a_1}g(t_f,x)f(t_f,x)\mathrm{d}x.
\end{align*}
From where we have got the adjoint operator $(\partial_{\mathbf{p}}\mathcal{P}(p,\mathbf{u}))^*$, which reads:
$$
(\partial_{\mathbf{p}}\mathcal{P}(p,\mathbf{u}))^*f=\frac{\partial f}{\partial t}-\gamma_x^1 x\partial_{x}  f+ \hat{I}(f)(t,x)- k_m^1  \! c_1(x;\mathbf{u})f+f(t_f,x).
$$
Moreover,  the definition of  $\mathcal{P}$ also includes boundary conditions, by deriving these boundary conditions we obtain that:
$$
 f(t,x)=0, \; \; (t,x) \in \Sigma.
$$

The multidimensional case can be obtained in the same way, applying the same method within the interval $[0,a_i[$ for $i=1,\dots,n$ and by summing these terms we get:
$$
(\partial_{\mathbf{p}}\mathcal{P}(p,\mathbf{u}))^*f= \dfrac{\partial f}{\partial t}(t,\mathbf{x})-\Gamma \cdot \nabla  f + \hat{I}(f)-\sum_{i=1}^{n} k_m^i c_i(x,\mathbf{u})f(t,x),
$$
where the multidimensional integral term is defined as 
$$
\hat{I}(f)=\sum_{i=1}^{n}\left(k_m^i c_i(\mathbf{x}_i;u)\int_{x_i}^{a_i} \! \omega_i(y_i-x_i)f(t,\mathbf{y}_i) \, \mathrm{d}y_i\right).
$$
Now, by taking into consideration that 
$$
N'(S(\mathbf{u}))=(S(u)(tf,x)-p_d(x))
$$
for $S(\mathbf{u})=p$, the equality \ref{adjointderiv} yields the following adjoint formulation:
\begin{equation}
    \mathcal{A}(f,u)=\begin{cases}
         -\dfrac{\partial f}{\partial t}(t,\mathbf{x})&=-\Gamma \cdot \nabla  f + \sum_{i=1}^{n}\left(k_m^i c_i(\mathbf{x}_i;u)\int_{x_i}^{a_i} \! \omega_i(y_i-x_i)f(t,\mathbf{y}_i) \, \mathrm{d}y_i\right) \\
         &-\sum_{i=1}^{n} k_m^i  \!
         c_i(x,u)f(t,x), \\
	f(t_f,x)&=-\left(p\left(t_{f}, \cdot \right)-p_{d}\left(\cdot \right)\right) \; \; x \in \Omega,\\
   f(t,x)&=0, \; \; (t,x) \in \Sigma. \\
    \end{cases}
    \label{SI_adjoint}
\end{equation}

\section{Further results}
\begin{figure}[H]
    \centering
    \includegraphics[scale=0.9]{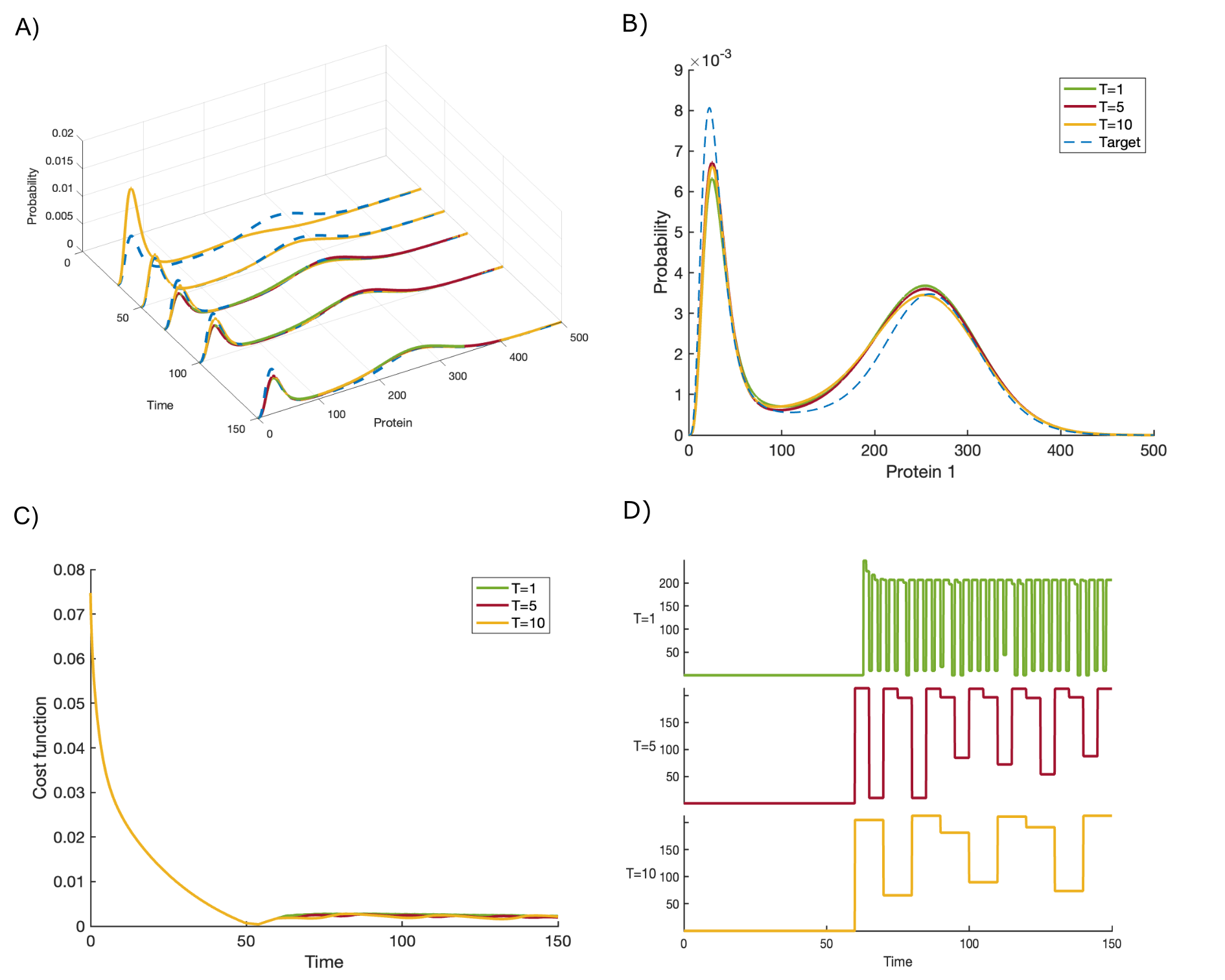}
    \caption{MPC results for the positvely auto regulated network with different sampling times $T=10,\; 20$ and 30 first mode $x_1=25$ second mode $x_2=250$ and final time $t_f=150$ A) MPC performance at different time points B) Final distributions for different sampling times C) Cost function evolution in time D) Input profiles for different sampling times  }
    \label{fig:bimodal_mpc_sampling}
\end{figure}

\begin{figure}[H]
\includegraphics[scale=0.9]{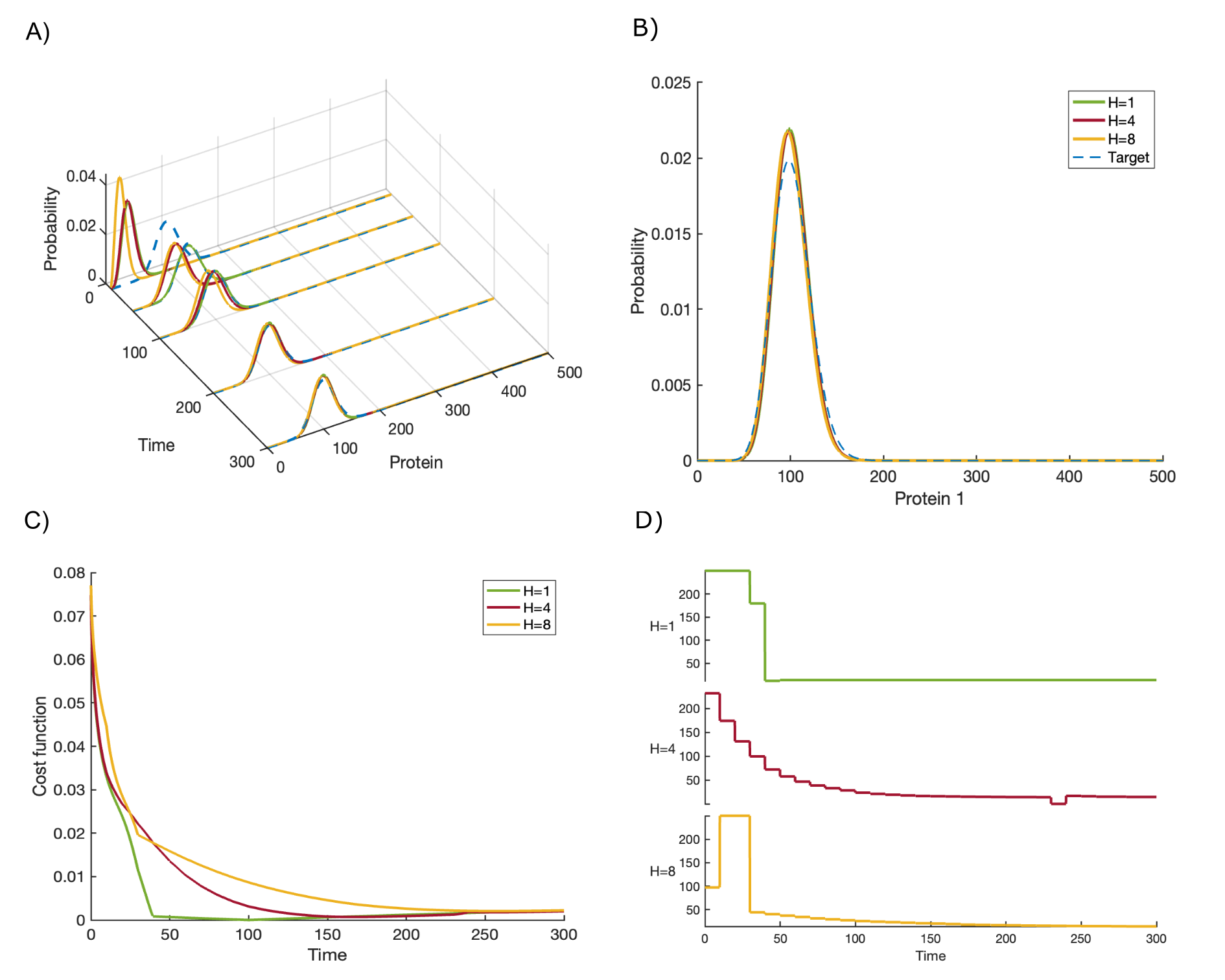}
  \caption{Results for different prediction horizons: A) Temporal evolution of the PIDE solution for different prediction horizons; B) mode simulations for different horizons at a $t_f=300$ min; C) cost function for each prediction horizon D) input profiles corresponding to the different prediction horizons. }
  \label{fig:opto_mpc_results_horizon}
\end{figure}

In this section we present further computational results related to the tow examples in the main text. Another important aspect we have considered is the effect of the sampling time. To investigate this effect on the MPC alrogithm described in the main text for the auto-regulatory network, we fix a prediction horizon on $H=1$, then we test the method for different $T=\{10 ~\textnormal{min},20 ~\textnormal{min},30 ~\textnormal{min}\}$  while keeping the same $P_{reach}$ as obtain in the reachability text. As it can be observed in Figure \ref{fig:bimodal_mpc_sampling}, there is not a major difference between the performance of the controller, and there is only some slight oscillations for larger sampling times. Regarding the reachability performance for the second example, the details are given in table \ref{tab:reachability_time_opto}, we also tested the performance of controller on the second example of the main paper under different prediction horizons $H=\{1,4,8\}$ by fixing  the sampling time to  10 (min) and the same target, as observed in the figure \ref{fig:opto_mpc_results_horizon} the larger the horizon the longer it takes to reach the target, but once the target is reached the performance is similar, once again this shows that the best result we obtained are with the shortest prediction horizon.

\begin{table}[h]
\begin{adjustwidth}{-2.25in}{0in} 
\centering
\caption{
{\bf Computational performance of the reachability under various conditions where test was run in the interval $T_f^0=100$ and $T_f^{max}=300$ (min), $\gamma$ is the variance of the target distribution}}
\begin{tabular}{|l|c|}
\hline
\multicolumn{1}{|l|}{\bf Mode 2} & 
\multicolumn{1}{|c|}{\bf Computation time (s)}\\ \thickhline
 200, $\gamma=15$ &   685 \\ 
 \hline
 80, $\gamma=15$ & 1053   \\
 \hline
30, $\gamma=15$ &   1108 \\
 \hline
 80, $\gamma=25$ &  793 \\
 \hline
 140, $\gamma=25 $ &  327 \\ 
 \hline
\end{tabular}
\label{tab:reachability_time_opto}
\end{adjustwidth}
\end{table}

\newpage
\appendix
\section{Notations}
\begin{itemize}
    \item[] $L^n(\Omega)$ for $1\leq n<\infty$ and $\Omega \subset \mathbb{R}^n$, is the space of Lebesgue measurable functions $f$ on $\Omega$ satisfying that $\displaystyle \int_\Omega |f(x)|^n \mathrm{d}x < \infty$.
    \item[] $H=L^2(\Omega)$. 
    \item[] $\mathcal{C}\left(\mathbb{R}_{+} ; L^1\left(\mathbb{R}^n\right)\right)$ is the space of continuous functions from $\mathbb{R_+}$ to $L^1\left(\mathbb{R}^n\right)$.
    \item[] $H^1(\Omega)=\{ f \in L^2(\Omega): \partial_{x_i} f \in L^2(\Omega)\; \; \text{for}\; \; i=1,\dots,n \}.$
    \item[] $ H^1_0(\Omega)=\{ f \in H^1(\Omega) : f(x)=0\; \; \text{on}\; \; \partial \Omega \}$. 
    \item[] $V =\{ f \in H^1(\Omega) : f(x)=0 \text{ on } \partial^+ \Omega \}$. 
     \item[] $F=\left\{f \in L^2(I ; V) : \partial_t f \in L^2(I ; V^*)\right\}$.   
\end{itemize}


\end{document}